\RequirePackage{fix-cm}
\documentclass[smallextended]{svjour3}       
\smartqed  
\usepackage{graphicx}
\usepackage{color}
\usepackage{amsmath}
\usepackage{natbib}
\usepackage{xcolor}
\usepackage{mdframed}
\usepackage{enumitem}
\usepackage{hyperref}
\setlist{nolistsep,leftmargin=*}
\usepackage{array}

\begin{document}

\title{Voyage through the Hidden Physics of the Cosmic Web
}


\author{Aurora Simionescu$^{1,2,3}$ \and
       Stefano Ettori$^{4,5}$ \and
       Norbert Werner$^{6}$ \and
       Daisuke Nagai$^{7,8}$ \and
       Franco Vazza$^{9}$ \and
       Hiroki Akamatsu$^1$ \and
       Ciro Pinto$^{10}$ \and
       Jelle de Plaa$^1$\and
       Nastasha Wijers$^2$ \and 
       Dylan Nelson$^{11}$ \and
       Etienne Pointecouteau$^{12}$ \and
       Gabriel W. Pratt$^{13}$ \and
       Daniele Spiga$^{14}$ \and
       Giuseppe Vacanti$^{15}$ \and
       Erwin Lau$^{16}$ \and
       Mariachiara Rossetti$^{17}$ \and
       Fabio Gastaldello$^{17}$ \and
       Veronica Biffi$^{18,19}$ \and
       Esra Bulbul$^{20}$ \and
       Maximilien J. Collon$^{15}$ \and
       Jan-Willem den Herder$^1$ \and
       Dominique Eckert$^{21}$ \and
       Filippo Fraternali$^{22}$ \and
       Beatriz Mingo$^{23}$ \and
       Giovanni Pareschi$^{14}$ \and
       Gabriele Pezzulli$^{24,22}$ \and
       Thomas H. Reiprich$^{25}$ \and
       Joop Schaye$^2$ \and
       Stephen A. Walker$^{26}$ \and
       Jessica Werk$^{27}$
}

\authorrunning{A. Simionescu et al.} 

\institute{
A. Simionescu \\
\email{a.simionescu@sron.nl} \\
$^1$ SRON Netherlands Institute for Space Research, Sorbonnelaan 2, 3584 CA Utrecht, The Netherlands\\
$^2$ Leiden Observatory, Leiden University, PO Box 9513, 2300 RA Leiden, The Netherlands \\
$^3$ Kavli Institute for the Physics and Mathematics of the Universe (WPI), The University of Tokyo, Kashiwa, Chiba 277-8583, Japan \\
$^4$ INAF, Osservatorio di Astrofisica e Scienza dello Spazio, via Pietro Gobetti 93/3, 40129 Bologna, Italy \\
$^5$ INFN, Sezione di Bologna, viale Berti Pichat 6/2, I-40127 Bologna, Italy \\
$^6$ Department of Theoretical Physics and Astrophysics, Masaryk University, Kotl\'a\v{r}sk\'a 2, 611 37 Brno, Czech Republic \\
$^7$ Department of Physics, Yale University, PO Box 208101, New Haven, CT 06520, USA \\
$^8$ Department of Astronomy, Yale University, PO Box 208101, New Haven, CT 06520, USA \\
$^{9}$ Dipartimento di Fisica e Astronomia, Universit\`a di Bologna, Via Gobetti 92/3, 40121, Bologna, Italy\\
$^{10}$ INAF - IASF Palermo, Via U. La Malfa 153, I-90146 Palermo, Italy \\
$^{11}$ Max-Planck-Institut f\"{u}r Astrophysik, Karl-Schwarzschild-Str. 1, 85741 Garching, Germany\\
$^{12}$ IRAP, CNRS, CNES, Universit\'e de Toulouse, France\\
$^{13}$ AIM, CEA, CNRS, Université Paris-Saclay, Université Paris Diderot, Sorbonne Paris Cité, F-91191 Gif-sur-Yvette, France\\
$^{14}$ INAF -- Osservatorio Astronomico di Brera, via Bianchi 46, 23087 Merate (LC), Italy \\
$^{15}$ cosine measurement systems, Oosteinde 36, 2361 HE Warmond, The Netherlands \\
$^{16}$ Department of Physics, University of Miami, Coral Gables, FL 33124, U.S.A. \\
$^{17}$ INAF-IASF Milano, via A. Corti 12, 20133, Milano, Italy \\
$^{18}$ Universit\"ats-Sternwarte M\"unchen, Fakult\"at f\"ur Physik, LMU Munich, Scheinerstr. 1, 81679 M\"unchen, Germany \\
$^{19}$ Harvard-Smithsonian Center for Astrophysics, Cambridge, MA 02138, USA \\
$^{20}$ Max Planck Institute for Extraterrestrial Physics, Giessenbachstrasse 1, D-85748 Garching bei M\"unchen, Germany\\
$^{21}$ Department of Astronomy, University of Geneva, ch. d’Ecogia 16, 1290 Versoix, Switzerland \\
$^{22}$ Kapteyn Astronomical Institute, University of Groningen, Landleven 12, 9747 AD Groningen, The Netherlands \\
$^{23}$ School of Physical Sciences, The Open University, Walton Hall, Milton Keynes, MK7 6AA, UK \\
$^{24}$ Department of Physics, ETH Zurich, Wolfgang-Pauli-Strasse 27, 8093 Zurich, Switzerland \\
$^{25}$ Argelander Institute for Astronomy, University of Bonn, Auf dem H\"ugel 71, 53121 Bonn, Germany \\
$^{26}$ Department of Physics and Astronomy, The University of Alabama in Huntsville, 301 Sparkman Drive NW, Huntsville, AL 35899, USA \\
$^{27}$ Department of Astronomy, University of Washington, Seattle, WA 98195, USA \\
}

\date{Received: date / Accepted: date}

\maketitle

\begin{abstract}
The majority of the ordinary matter in the local Universe has been heated by strong structure formation shocks and resides in a largely unexplored hot, diffuse, X-ray emitting plasma that permeates the halos of galaxies, galaxy groups and clusters, and the cosmic web. We propose a next-generation ``Cosmic Web Explorer" that will permit a complete and exhaustive understanding of these unseen baryons. This will be the first mission capable to reach the accretion shocks located several times farther than the virial radii of galaxy clusters, and reveal the out-of-equilibrium parts of the intra-cluster medium which are live witnesses to the physics of cosmic accretion. It will also enable a view of the thermodynamics, kinematics, and chemical composition of the circumgalactic medium in galaxies with masses similar to the Milky Way, at the same level of detail that $Athena$ will unravel for the virialized regions of massive galaxy clusters, delivering a transformative understanding of the evolution of those galaxies in which most of the stars and metals in the Universe were formed. Finally, the proposed X-ray satellite will connect the dots of the large-scale structure by mapping, at high spectral resolution, as much as 100\% of the diffuse gas hotter than $10^6$ K that fills the filaments of the cosmic web at low redshifts, down to an over-density of 1, both in emission and in absorption against the ubiquitous cosmic X-ray background, surveying at least 1600 square degrees over 5 years in orbit. This requires a large effective area ($\sim$10 m$^2$ at 1 keV) over a large field of view ($\sim1$ deg$^2$), a megapixel cryogenic microcalorimeter array providing integral field spectroscopy with a resolving power $E/\Delta E$ = 2000 at 0.6 keV and a spatial resolution of 5$^{\prime\prime}$ in the soft X-ray band, and a low and stable instrumental background ensuring high sensitivity to faint, extended emission.
\end{abstract}

\section{Diffuse matter in the post-\textit{Athena} era}

Most of the Universe is invisible: 95\% of its contents consist of dark matter and dark energy, which we do not yet understand. But even when it comes to the ``normal'' standard-model particles, we can only see the tip \textit{of the tip} of the iceberg. A large fraction of the baryons have not been converted into stars, but instead reside in the hot, diffuse medium that fills extended galaxy halos, galaxy groups, galaxy clusters, and the cosmic web. These environments are best probed by observations at soft X-ray wavelengths ($\sim 10-100$ \AA ), requiring spaceborne observatories. 

The majority of X-ray observations so far have naturally focused on the densest, brightest centers of clusters and groups of galaxies, revealing in detail the physics of only a tiny fraction of the hot, diffuse matter that permeates the Universe. Even there, after 20 years of exquisite observations and discoveries with \textit{Chandra} and \textit{XMM-Newton}, many questions still loom. High-resolution X-ray spectroscopy studies of the intra-cluster medium (ICM) are all but lacking, leaving a huge gap in our knowledge of the dynamical nature of this hot, diffuse plasma. The \textit{Athena} observatory is set to revolutionize this field, and significantly advance our understanding of the ``Hot Universe''.

\begin{mdframed}[backgroundcolor=blue!20] 
To further reveal how the cosmic web is interconnected, we must \textbf{survey and physically characterize the vast majority of the very faint warm-hot diffuse baryons in the local Universe}. This poses unique challenges that no existing or planned telescope has been designed to address thus far. 
\end{mdframed}

\noindent\textbf{What are we still missing?} 
\begin{enumerate}
\item All of the X-ray instruments approved so far only aim to measure the properties of the ICM in massive galaxy clusters within a limited radial range, typically up to $r_{200c}$ \footnote{the radius within which the mean enclosed density is 200 times the critical density at the redshift of the cluster.}. Mapping the physics, kinematics, and chemistry within the \textit{entire} hot gaseous halo of a single, massive, $M_{\mathrm{virial}}\sim10^{15}M_\odot$, z=0.1 galaxy cluster, expected to extend 4--5 times farther than $r_{200c}$ and thus cover more than 5 deg$^2$ on the sky, would require a whopping mosaic of 1000 pointings with the \textit{Athena} X-IFU, most of these with an exposure time well in excess of 1~Ms (meaning a total observing time of over 30 years). \textbf{The most exciting, out-of-equilibrium parts of galaxy clusters, located beyond the virial radius and which are live witnesses to the physics of cosmic accretion, would remain entirely unexplored} in the absence of a new X-ray mission with a significantly larger grasp.  
\item Massive, X-ray bright clusters of galaxies are rare, and represent only a small fraction of the matter in the Universe. The far less massive, and far fainter, \textbf{soft X-ray emitting halos of $\sim$L$^\star$ galaxies are poorly understood, although it is in these halos that most of the stars and metals in the Universe were formed}. The dominant emission from these lower-mass halos are in the OVII and OVIII multiplets, where the resolving power of \textit{Athena}'s X-IFU, while excellent and unprecedented at higher energies, is only R$\sim$300. This is insufficient to measure typical velocities of $\sim100$ km/s expected to be associated with the cycling of baryons through the circumgalactic medium (CGM). We need a future X-ray mission that will revolutionize the studies of the CGM in galaxies with masses similar to that of the Milky Way, in the same way that \textit{Athena} will revolutionize studies of clusters of galaxies. 
\item The diffuse matter permeating large-scale structure (LSS) filaments remains elusive. \textit{Athena} will allow a first systematic study of this so-called Warm-Hot Intergalactic Medium (WHIM), by detecting it in absorption along 200 sightlines towards bright BL Lacs and gamma-ray bursts (GRB), and studying its corresponding emission spectrum in a handful of cases. However, these observations are contingent upon the chance existence of bright (and thus rare) background beacons to illuminate the WHIM, and will only probe its properties along sparse and narrow pencil-beam sight lines. Obtaining \textbf{a complete 3D picture of the baryons permeating the spatially complex large-scale structure requires wide-field, very sensitive tomographic observations of soft X-ray emission}, in combination with absorption studies that can make use of much fainter background sources offering a more uniform sky coverage. 
\end{enumerate}

\noindent The CGM, cluster outskirts, and WHIM are intimately interrelated. 
Like blood circulating through the human body, the chemical elements produced in stars, pumped by the energy from supernovae (SNe) and supermassive black holes (SMBH), cycle through the Universe's large-scale structure. Metals often escape the shallow gravitational potential wells of the galaxies where they were produced; from there, they either get re-accreted into the CGM, or become mixed into the diffuse LSS filaments and are then funneled into the outskirts of galaxy clusters, the most massive knots of the cosmic web. To really connect the dots of the large-scale structure and to understand this circulation in detail we need a \textbf{\textcolor{purple}{Cosmic Web Explorer}} that will reach unprecedented X-ray sensitivity limits over unprecedented areas on the sky. Beyond a much larger mirror collecting area and field of view (FoV), a much lower and more stable instrumental background, and an improved spectral resolution at the OVII and OVIII lines, this also requires a very accurate understanding of the X-ray halo of our own Milky Way which acts as a foreground to the faint emission we are searching for. 
{\bf Only a next-generation mission that will survey a large area of the sky using sensitive, high spatial and spectral resolution integral field spectroscopy in the soft X-ray band can fully achieve this goal, building upon previous progress brought about by \textit{XRISM}\footnote{https://global.jaxa.jp/projects/sas/xrism/} and \textit{Athena}\footnote{https://www.the-athena-x-ray-observatory.eu/}.}

\section{The unknowns of the unseen cosmic web in X-ray light}

\subsection{The emergent large-scale structures around the knots of the cosmic web} 
\label{sect:cluster}

\begin{mdframed}[backgroundcolor=blue!20] 
Galaxy clusters are the ultimate manifestation of hierarchical structure formation, and they continue to grow and accrete matter at the present time. The outer regions of galaxy clusters are home to the majority of the diffuse gas in these systems, and bear witness to the complex physics of large-scale structure growth as it happens. 
A plethora of unexplored structure formation physics is believed to be operating near and beyond the virial radii of galaxy clusters, and these processes are fundamentally different from the physics in the cores of clusters that has been the focus of X-ray cluster science over the past several decades.
\end{mdframed}

An ultimate census of the baryons, even inside massive clusters of galaxies, can only be achieved by (1) mapping the entire volume of clusters in order to identify and characterize substructures on both small and large scales, and (2) accounting for bulk and turbulent gas motions, unresolved clumping, and non-equilibrium phenomena that would otherwise significantly bias the gas density, temperature, metal, and mass measurements. 
\textbf{The rich thermal, kinematic, and chemical contents of cluster outskirts are the Rosetta stone for understanding the growth of galaxy clusters and their connections to the Cosmic Web, as well as a stepping stone towards exploring the outskirts of massive galaxies and galaxy groups.} 

\begin{figure}[!ht]
\centering
\includegraphics[width=\textwidth]{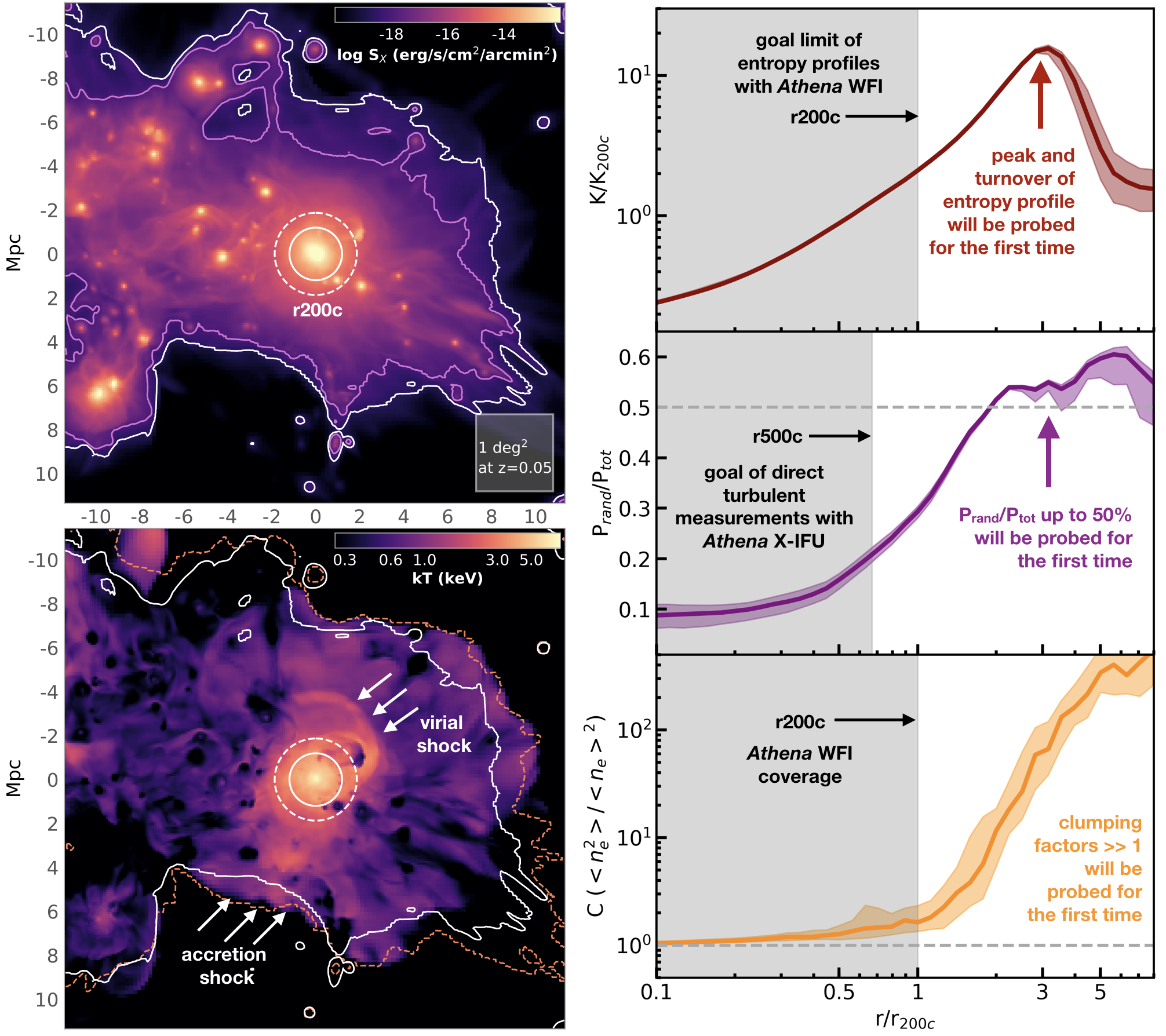}
\caption{\footnotesize{Example of a massive, relatively relaxed cluster from the {\em Omega500} adiabatic cosmological simulation \citep{nelson14}. Top left: predicted X-ray emissivity in the 0.5--2 keV band. Purple contours show $S_X > 10^{-18} {\rm erg\, s^{-1}\,cm^{-2}\,arcmin^{-2}}$, corresponding to the spectral simulations in Figure \ref{fig:emis_spectra}. Assuming an accurate understanding of the cosmic foregrounds and backgrounds, in an exposure time of 1~Ms the proposed mission can reach $S_X > 2\times10^{-19}\: {\rm erg\, s^{-1}\,cm^{-2}\,arcmin^{-2}}$ (white contours) over an extraction region of 1000 arcmin$^2$ at the 5$\sigma$ level.
Bottom left: projected temperature. Surface brightness contours at $S_X > 2\times10^{-19}\: {\rm erg\, s^{-1}\,cm^{-2}\,arcmin^{-2}}$ (white) and for a column density of $N_{\rm H} > 8 \times 10^{18}$ cm$^{-2}$ (orange; to be probed in absorption against any bright quasars in the FoV) are shown. Solid and dashed circles show $r_{500c}$ and $r_{200c}$, respectively. 
Right panel: predicted radial profiles of the entropy, ratio of the turbulent to total pressure, and gas density clumping factor in the ICM for the {\em Omega500} cluster sample, illustrating that the observations proposed here will probe a new regime of virialization and deviations from equilibrium that has never been reached before.}}
\label{fig:cluster_sx}
\end{figure}

\subsubsection{The shocked baryons at the edge of galaxy clusters}

The outermost boundary of the X-ray emitting gas halo of galaxy clusters is marked by the so-called ``accretion shock'' or ``external shock''. 
It is here, around 4--5~$r_{200c}$, that low-temperature, low-density gas accreting from the void regions is heated by strong shocks with Mach numbers of several tens to hundreds, reaching X-ray emitting temperatures during its first infall into the cluster potential \citep[e.g.][]{ry03,molnar09}. ``Internal shocks'' or ``virial shocks'' due to mergers and filamentary accretion further increase the entropy of the gas. \textbf{Although directly responsible for heating most of the baryons in the Universe into a hot and diffuse state, neither ``virial shocks'' nor ``accretion shocks'' have ever been probed observationally.} 

\begin{figure*}[tb]
\centering
\includegraphics[width=\textwidth]{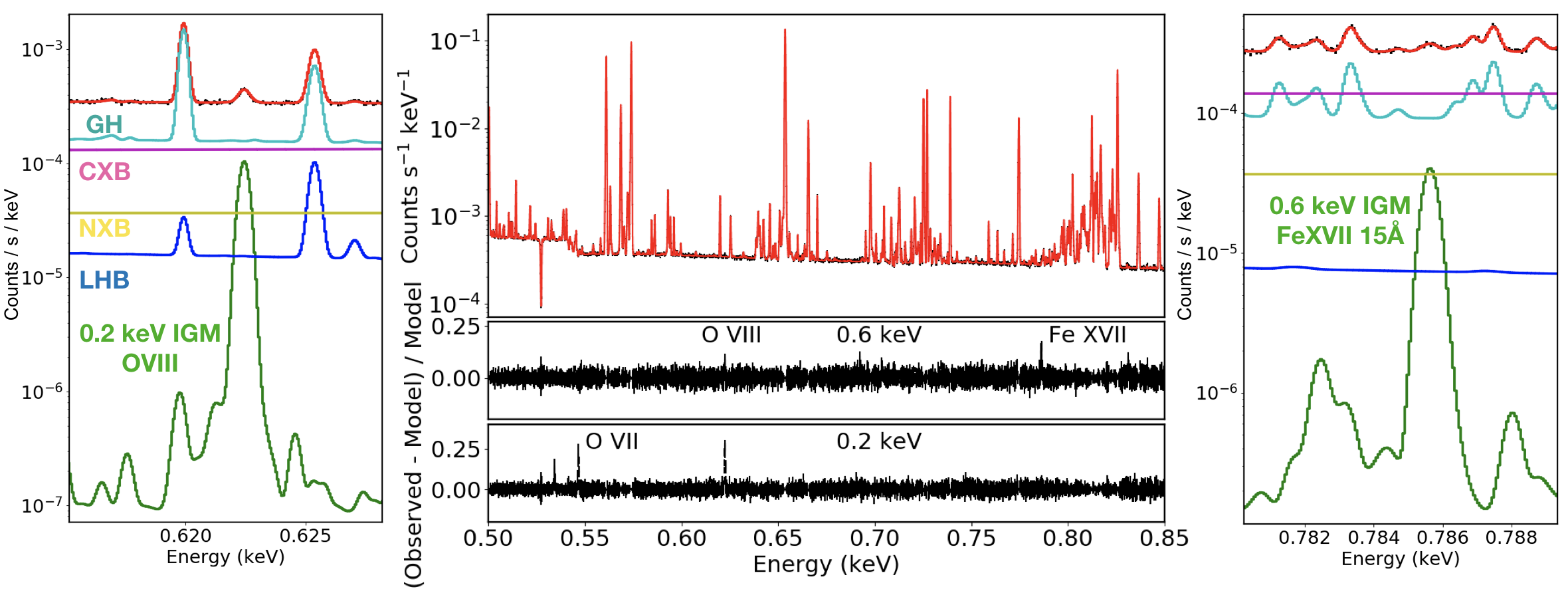}
\caption{\footnotesize{Simulations of faint, diffuse emission with a source surface brightness of $10^{-18}$ erg/s/cm$^2$/arcmin$^2$ in the 0.5--2 keV band, metallicity of 0.3 Solar, and turbulent broadening of 100~km/s. We include the Galactic halo (GH), local hot bubble (LHB), and cosmic and instrumental X-ray backgrounds (CXB and NXB), and assume an exposure time of 1~Ms with the proposed mission configuration, and an extraction area of 1000 arcmin$^2$. Residual plots in the bottom middle panels show that several emission lines will be significantly detected for a range of temperatures typical inside the cluster accretion shock (0.6~keV, Fig. \ref{fig:cluster_sx}) and denser parts of the WHIM (0.2~keV). By covering a large sky area with high sensitivity, the proposed mission will uniquely allow us to map and model the emission from the Galaxy and unresolved AGN at the sub-percent level accuracy required for a robust characterization of the source signal.}}
\label{fig:emis_spectra}
\end{figure*}

\textbf{\textcolor{purple}{How did the hot Universe become hot?}} To really understand the process of virialization and heating of the ICM, we need direct measurements probing the peak and turnover of the entropy profile at and beyond the virial radius of the cluster (as shown in the top-right panel in Figure \ref{fig:cluster_sx}). For a massive, $M_{\mathrm{virial}}\sim10^{15}\:M_\odot$ cluster, this requires reaching surface brightness levels as faint as $S_X<10^{-18}$ erg/s/cm$^2$/arcmin$^2$ in the 0.5--2 keV energy band, for temperatures around $kT\sim0.6$ keV (with lower $S_X$ and $kT$ for lower-mass halos). Figure \ref{fig:emis_spectra} shows the feasibility and challenges of reaching such faint flux levels.  

\subsubsection{Unveiling the rain and streams of plasma accreted from the surrounding LSS}\label{sect:clump}

Both numerical simulations and observations show that deviations from a smooth, spherical distribution become increasingly important towards the outer edges of clusters of galaxies, with inhomogeneities manifesting themselves over a broad range of spatial scales (for a recent review, see \citealt{Walker2019}). On large scales, high density, low entropy streams of gas from cosmic web filaments, coherent over mega-parsec scales, are predicted to penetrate deep into the cluster interior, generating bulk and turbulent gas motions, and producing shocks and contact discontinuities as they interact with the surrounding, virialized ICM \citep{Zinger2016,Zinger2018}. On smaller scales, infalling gas substructures around tens of kpc across lead to gas clumping that is ubiquitous throughout the cluster outskirts \citep[e.g.][]{nagai11,zhuravleva13,roncarelli13,vazza13,battaglia15}. The contributions of these infalling clumps to X-ray emission increases toward the low-density region in cluster outskirts, where they are less efficiently disrupted by ram-pressure stripping from the surrounding ICM. At present, largely due to their low surface brightness, the physical properties of these gas clumps and streams remain almost completely unexplored.

\textbf{\textcolor{purple}{How is matter funneled into the most massive knots of the cosmic web?} How and when does the accreted matter mix with the rest of the ICM?} All existing major X-ray telescopes have, by now, dedicated extensive amounts of exposure time to understanding the thermodynamical properties of the ICM in the cluster outskirts, including \textit{Suzaku} \citep{urban14,simionescu13,simionescu17}, \textit{Chandra} \citep{morandi14}, and combinations of \textit{XMM-Newton} and Sunyaev-Zel'dovich (SZ) measurements with \textit{Planck} \citep{ghi19,ett19}. These observations have given us a first taste of the richness of cluster outskirts physics, revealing the onset of an increasingly inhomogeneous gas density \citep{Simionescu11,Eckert2012,Eckert15,Tchernin16}, and multiple large-scale structure filaments connecting massive clusters to the cosmic web (\citealt{werner2008,Eckert15_Nature,connor2018}; and, most recently, the deep scan observations of Abell 3391/95 obtained during the performance verification phase of \textit{eROSITA}; \citealt{reiprich2020}). 

But this is just the beginning. \textbf{Numerical simulations predict that the signatures of gas clumping become more and more dominant as we move beyond $r_{\rm 200c}$, into a radial regime that has yet to be probed routinely by X-ray observations (as shown in the bottom-right panel of Figure \ref{fig:cluster_sx}).}  
The Cosmic Web Explorer proposed here will map the X-ray emissivity in galaxy clusters $3-5$ times further in radius than the \textit{Athena} Wide Field Imager (WFI), out to the edge of their X-ray halos marked by the accretion shock. With its large grasp (16 times that of the WFI), a lower and more stable instrumental background, and maintaining a sufficient spatial resolution (5$^{\prime\prime}$) to identify both small-scale and large-scale asymmetries, \textbf{the proposed mission will reveal the full picture of large-scale structure formation that is currently hidden from us.}

\subsubsection {Turbulence and non-thermal pressure in the regions of ongoing virialization} 

The continuous accretion of gas and dark matter from the Cosmic Web is expected to convert a non-negligible fraction of the infall kinetic energy into the injection of \textit{non-thermal} energy across a wide range of scales. Recent simulations and observations indicate that the bulk of the non-thermal energy resides in subsonic chaotic gas motions in the ICM. These gas motions provide non-thermal pressure support against gravity, supplemental to thermal pressure. They are also a source of heat: kinetic energy is transported to smaller scales via progressively smaller vortices, and converted into thermal energy at the dissipation scale ($< 1$ kpc) \citep[e.g.,][]{gaspari14,zuhone16,shi18}. 

To date, X-ray and SZ observations have provided early, indirect evidence for the non-thermal pressure due to bulk and turbulent gas motions in nearby clusters, from their cores \citep{zhuravleva14} out to intermediate and large radii \citep{Khatri2016,eckert17,Siegel2018,eckert19}. 
The best direct measurements of bulk and turbulent gas velocities in the ICM to date have been obtained for the core of the Perseus cluster, where the \textit{Hitomi} X-ray observatory provided constraints on the Doppler shifting and broadening of the 6.7 keV Fe XXV K$\alpha$ emission line with a spectral resolution of $\sim$5~eV \citep{hitomi18}. This led to important physical insights into the nature of gas motions driven by mergers and accretion, and feedback from the active galactic nucleus (AGN) \citep{lau17,bourne17}. 
In the coming years, \textit{XRISM}/Resolve and the \textit{Athena}/X-IFU will extend measurements of the bulk and turbulent motions in the ICM to many other nearby galaxy clusters, to intermediate cluster radii ($<0.6 r_{200c}$), and to the cores ($<0.1 r_{200c}$) of higher redshift clusters. 
This will allow us to characterize the hydrostatic mass bias due to non-thermal pressure \citep{ota18}, and calibrate galaxy clusters as exquisite probes for precision cosmology; the ICM velocity power spectrum will be measured with \textit{Athena} down to several kpc scales, revealing ICM kinematics close to the dissipation scale in nearby clusters \citep[][]{roncarelli18,cucchetti19}.

Numerical simulations show that the fraction of non-thermal pressure due to gas motions increases with distance from the cluster centers \citep[e.g.,][]{lau09,vazza18}, and can become comparable to the thermal gas pressure at distances ($>r_{200c}$) that remain out of reach for all currently planned and proposed X-ray observatories (shown in the middle right panel of Figure \ref{fig:cluster_sx}). The non-thermal pressure fraction is also a sensitive indicator of the mass accretion rate of dark matter halos \citep[e.g.,][]{nelson14,shi15}.

By combining exquisite sensitivity to faint, diffuse emission, a large FoV, and sufficient spectral resolution to detect a turbulent broadening of $\sim 100$ km/s for the low-energy emission lines characteristic of cluster outskirts (e.g. FeXVII, see Fig. \ref{fig:emis_spectra}), the Cosmic Web Explorer is equipped with the ideal capabilities to answer questions like: \textbf{How does the ongoing accretion shape the kinematic properties of the ICM near and beyond the cluster virial radii? \color{purple} How can we characterize the physics of the gas far out of hydrostatic equilibrium, with comparable thermal and turbulent pressures?  \color{black} To what extent is the dissipation of turbulence a potential heating source that supplements the virial shocks? \color{black}} 

\subsubsection{The chemical enrichment recipe for the diffuse intergalactic baryons} 

Supernovae and supermassive black holes drive galactic winds, spreading the metals produced in stars into the surrounding intergalactic medium. By measuring how far these metals are spread, how many metals escape the halo of their host galaxy and when this process occurs, and by determining the relative chemical composition between various light and heavy elements, we obtain powerful probes of the properties of stellar- and AGN-driven winds, the integrated star formation history, and the physics of supernova explosions. 

The ICM offers an exceptionally clean probe to measure the chemical evolution of the Universe as a whole (for a recent review, see \citealt{mernier2018c}). Since the ICM plasma is well approximated as optically thin and in collisional ionization equilibrium, the equivalent widths of detected emission lines can be easily converted into elemental abundances. A relatively narrow energy band spanning 0.3--3 keV is expected to contain a wealth of line emission from elements between C and Ni. 

In particular, the metallicity distribution in the outskirts of galaxy clusters is emerging as an important test bed of feedback physics, wherein the uniform level of chemical enrichment observed throughout the outer regions of massive clusters \citep{Werner2013,Urban2017} requires a significant injection of metals from SMBH at early times \citep{Biffi2018}. The relative composition of various elements of the ICM also places constraints on the star formation histories and the chemical evolution of the Universe \citep{Simionescu2015}; such constraints are especially interesting at the periphery of clusters and beyond, where the outskirts connect to cosmic filaments.

Measurements of metallicity in cluster outskirts are extremely challenging, and only exist for a handful of very nearby, bright, galaxy clusters. An excellent spectral resolution for diffuse sources is necessary for a reliable determination of the equivalent widths of faint emission lines \citep[as demonstrated by][]{Hitomi2017Z}. While the spectrometers on \textit{XRISM}/Resolve and the \textit{Athena}/X-IFU will undoubtedly reveal invaluable information regarding the chemical enrichment pattern in the cores of galaxy clusters and groups across cosmic time, the combination of high-resolution spectroscopy and a very large grasp offered by the proposed 
Cosmic Web Explorer is required to probe the metal abundance ratios in the outskirts of these halos, which span many square degrees on the sky, and give a definitive answer to the question: \textbf{\color{purple} what is the recipe for distributing the building blocks of life throughout the bulk of intergalactic space?} \color{black} As shown in Figure \ref{fig:emis_spectra}, lines from both O (predominantly contributed by core-collapse supernovae) and Fe (primarily a SN~Ia product) can be detected even for surface brightness levels expected far beyond $r_{200c}$, allowing us to determine how different enrichment sources contributed to the metal budget the region of ongoing virialization at the edge of the clusters' X-ray halos. 

\subsubsection{Non-equilibrium phenomena in cluster outskirts}

Besides thermal, kinematic, and chemical properties of the ICM, the cluster formation process gives rise to a variety of still poorly understood \textit{plasma physics} across a wide range of scales. For example, in the extremely low-density regions in cluster outskirts, the Coulomb collision time of electrons and protons becomes longer than the age of the Universe \citep[e.g.][]{rudd09,avestruz15}. \textbf{How do electrons get heated in cluster outskirts? \textcolor{purple}{What is the role of magnetic fields in mediating the equilibration between different particle species in the plasma?} Does the ideal fluid approximation, which is often employed in numerical simulations of large-scale structure formation, break down? If so, at what point?}

\textit{Hitomi} offered us a first glimpse into the power of high-resolution spectroscopy to probe deviations from the collisional ionization approximation \citep{hitomi2018t}, and measure the ion temperature independently of the electron temperature \citep{hitomi18} in the bright core of the Perseus Cluster. These diagnostic tools require excellent spectral resolution and a very large number of spectral counts which, for the faint and very extended cluster outskirts, can only be obtained with the proposed Cosmic Web Explorer. 

Detecting the signature of non-thermal phenomena in cluster outskirts has the potential to provide a view of out-of-equilibrium plasma conditions, where relativistic particles can be accelerated in a so-far unexplored plasma regime \citep[e.g.][]{2019SSRv..215...14B}
and give rise to observed diffuse radio emission by interacting with diffuse magnetic fields \citep[e.g.][]{bj14}. Key unknown quantities to understand how these processes proceed are the Mach number of accretion shocks, the exact thermodynamic structure of post-shock relaxation regions, as well as the degree of plasma collisionality here \citep[e.g.][]{by08,bl11b} -- quantities which the proposed mission is ideally equipped to probe. Combined with forthcoming constraints on the non-thermal phenomena of the Universe from radio (e.g. LOFAR, MWA, ASKAP, MEERKAT, and SKA; see discussion in \S~\ref{sec:synergies}), \textbf{the Cosmic Web Explorer will allow us to connect the physics of the thermal and relativistic large-scale Universe in unprecedented detail.}

Fundamental multi-scale plasma-physics questions related to the heating, acceleration, and partitioning of energy between electrons and ions, as well as the role of magnetic fields and discontinuous processes such as shocks and reconnection, underpin the topics addressed in this and many other White Papers included in this special issue. These common questions are best approached by combining cross-disciplinary studies covering a very wide range of plasma conditions. The Cosmic Web Explorer represents an important and unique addition to the parameter space that can be probed using other plasma structures such as cometary tails, the Sun, or in-situ measurements of the near-Earth plasma environment.

\subsection{The circumgalactic medium as a driver of galaxy evolution}
\label{sect:cgm}

Most of the stars and metals in the Universe were formed in approximately Milky Way mass,  $M_{\rm tot}\approx10^{12}~M_\odot$, galaxies. The majority of baryons in these $L^\star$ galaxies reside in shock heated atmospheres with temperatures of millions of degrees, extending far beyond the stellar component. About half of the yet unseen warm-hot diffuse matter in the local Universe may lie in such extended galactic atmospheres \citep[e.g.][]{fukugita1998,keres2005,fukugita2006}.
\textbf{Gaseous halos are inextricably linked to their host galaxies through a complex story of accretion, feedback, and continual recycling.} The energetic processes that define the state of gas in the CGM are the same ones that regulate stellar growth and create the diversity of today's galaxy colors, star formation rates, and morphologies, spanning Hubble's Tuning Fork Diagram. 

\textbf{\color{purple}
How does matter cycle between galaxies and structures on larger scales?} Our understanding of galaxy evolution is critically limited by our poor understanding of the cycling of baryons through the circumgalactic and intergalactic media.
We are in the era when UV absorption studies, mainly focused on the OVI emission line doublet, but also probing many other metal ions with ionization potential energies $<$ 10 Ryd, are dramatically increasing our knowledge of the CGM \citep[e.g.][]{tumlinson2011,Werk:2016aa}. These observations are shifting focus to the CGM as one of the most crucial probes of galaxy evolution. 
Despite the large interest generated by these observations in the community, we are still unclear about the nature of the absorbing gas: is it hot collisionally ionized gas \citep{oppenheimer2016}, warm photo-ionized gas \citep{stern2016,oppenheimer2018a}, or conductive layers at the interface between different media \citep{armillotta2017}? Determining which of these scenarios is correct would have important implications for our understanding of the multiphase nature of the CGM and the cycle of baryons around galaxies \citep{McQuinn:2018aa}. This goal, however, cannot be achieved based only on the pencil-beam views offered by absorption studies and, most importantly, without the more comprehensive view based on the study of higher ionization states of oxygen and other metals.

\begin{figure}[t]
\centering
\includegraphics[width=\textwidth]{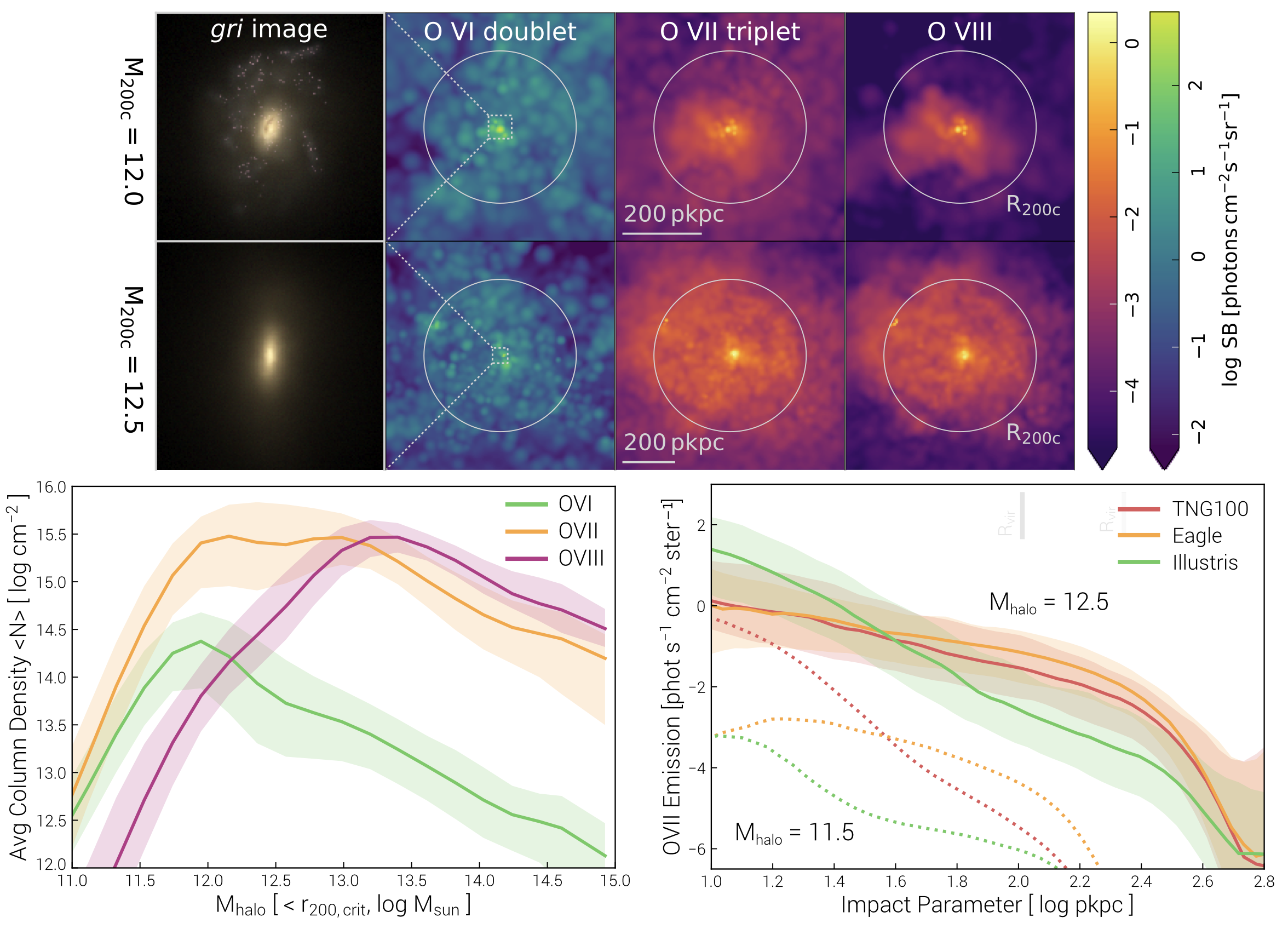}
\caption{\footnotesize Emission signatures from diffuse baryons in the CGM as predicted by three current large-volume cosmological hydrodynamical simulations: Illustris \citep{vogelsberger2014}, EAGLE \citep{schaye15}, and IllustrisTNG \citep{pillepich2018tng}. These three simulations invoke different galaxy formation physics and include a diverse range of black hole feedback models. (Top:) Predicted surface brightness maps in the OVI, OVII, and OVIII emission lines for two halos of different mass chosen from the EAGLE simulations, and their connection to the central galaxy at the heart of each halo. (Lower left:) OVII and OVIII ions dominate in abundance over the commonly observed UV-wavelength OVI ion, even in the CGM of Milky Way mass galaxies \citep{Nelson:2018aa}. (Lower right:) Different models predict a diversity of signatures for the thermal, kinetic, and ionization properties of the hot baryonic phase, particularly for lower-mass Milky Way and group-size halos, regimes largely unconstrained by current X-ray observational facilities.}
\label{fig:cgm}
\end{figure}

\begin{mdframed}[backgroundcolor=blue!20] 
The bulk of the gas-phase oxygen in the CGM is expected to be in the form of OVII and OVIII (Figure \ref{fig:cgm}, lower left panel), which both have strong soft X-ray emission line multiplets. 
Modern simulations which are able to reproduce the basic optical galaxy properties (Figure \ref{fig:cgm}, top panel), such as Illustris, IllustrisTNG, and EAGLE, differ in their predictions of the OVII and OVIII radial profiles by many orders of magnitude, particularly for lower-mass galaxies (below $\approx 10^{12}~M_{\odot}$, Figure \ref{fig:cgm}, lower right). Direct mapping of the intensity, as well as line of sight velocities and velocity dispersions of these lines will provide the robust and comprehensive understanding of the CGM required to constrain models of galaxy formation and evolution.
\end{mdframed}

The hot atmosphere of our own Milky Way Galaxy contains about  $2.5\times10^{10}~M_{\odot}$ of gas \citep{Bland-Hawthorn2016}, and the baryonic mass fraction within the virial radius is only $\approx6$\%, which falls well short of the Universal cosmic value of 16\%. Some kind of violent activity likely caused our Galaxy to lose a large part of its hot atmosphere. 

The Galactic centre shows abundant evidence for such violent activity: gamma-ray observations have identified giant cavities filled with relativistic plasma -- called Fermi bubbles -- on both sides of the Galactic plane, which indicate that our Galactic centre has recently released large amounts of energy. Recently, \citet{ponti2019} reported the discovery of prominent X-ray structures above and below the plane. These ``Galactic Centre Chimneys'' may constitute a channel through which energy and mass, injected at the Galactic centre, are transported to the base of the Fermi bubbles. These events may lead to the heating, enrichment, but also to the partial loss of the Galactic atmosphere. High resolution X-ray spectroscopy with \textit{XRISM} and \textit{Athena} will significantly improve our knowledge of the temperature structure and metallicity of Milky Way's atmosphere. However, the spectral resolution at low energies (corresponding to $>1000$ km/s for the X-IFU) will not be sufficient to constrain the velocity structure, and the small FoV of these instruments will only allow us to obtain pieces of the puzzle - not a contiguous map over a wide field.
{\bf An X-ray observatory with a large field of view and high spectral resolution in the soft band ($R=2000$ at 0.6 keV) will revolutionize how we view the dynamical interaction between the hot atmosphere of our own Galaxy, outflows and jets from the supermassive black hole SgrA$^{\star}$ at its center, and stellar feedback.}

Studies of other spiral galaxies with \textit{Chandra} and \textit{XMM-Newton} also find less baryonic mass in their atmospheres than expected. When the observed density profiles are extrapolated to the virial radius of these galaxies, more than 60--70\% of the baryons appear to be missing \citep{anderson2016,bogdan2017,li2018}. The atmospheres of at least some of these disky galaxies could be in an outflow state \citep{pellegrini2018}. \citet{bregman2018} show that the atmospheric density profiles of massive spirals need to be extrapolated out to 1.9--3 $r_{200c}$ for their baryon to dark matter ratio to approach the cosmic value. Even with long, $\sim$Ms observations, \textit{Athena} will only probe the central regions of $\sim L^{\star}$ galaxies \citep[out to 0.4 $R_{\rm virial}$,][]{kaastra2013}, and can not resolve their line widths and shifts, demonstrating the importance of a future instrument with much higher throughput and improved spectral resolution at soft X-ray energies.

The question of the baryon content of galactic atmospheres is further complicated by our lack of knowledge about their metallicity. Today, we only have reliable constraints on the chemical composition of hot atmospheres surrounding massive ellipticals, which have metallicities approaching the Solar value \citep{mernier2018}. In general, their metal abundances peak in their centres and flatten out at $\approx0.2-0.3$~Solar at larger radii \citep{mernier2018c}. Whether the same is true for lower-mass galaxies is still unknown. This is particularly important because these low-mass halos are increasingly dominated by stellar feedback, of which metals are faithful tracers.   

The CGM provides the fuel for galaxy formation and is fed by both cosmological accretion and feedback processes. 
$L^\star$ spiral galaxies that are forming stars at the rate of our Milky Way ($\approx2~M_\odot$~yr$^{-1}$) would consume the available molecular gas in their discs in about $10^9$ years. To maintain their star-forming galactic discs, they have to be fed continuously by molecular gas from outside \citep{fraternali2012}. Thermal instabilities in their hot atmospheres, leading to a ``rain'' onto the galactic disc, would be able to provide plenty of fuel to maintain the star formation in spirals for $\approx 10^{10}$ years.
During this phase of galaxy evolution, the hot atmosphere is not only a source of mass for the galaxy, but necessarily also of angular momentum: otherwise, galaxy discs would inevitably shrink with time, opposite to observations \citep{vanderWel2014}. While this important aspect has been the subject of recent theoretical studies \citep{pezzulli2017,oppenheimer2018b}, a robust observational determination of the rotation of hot atmospheres is close to impossible with the current instrumentation \citep{hodgeskluck2016}, further highlighting the need for the high spectral resolution mission proposed here.

Thermally unstable cooling from the hot galactic atmospheres is also relevant in early-type galaxies where it likely feeds the central supermassive black hole. When gas from the hot halo condenses into cooler clouds that ``precipitate'' toward the center \citep{gaspari2013,voit2015c,mcnamara2016}, the accretion rate can rise by orders of magnitude, triggering a feedback response by the AGN which heats the gas and balances its cooling. This closes the feedback loop needed to maintain a very delicate equilibrium, regulating the star formation rate and ensuring the co-evolution of its SMBH and host galaxy.

The detailed physics of the development of cooling instabilities and multi-phase gas, which is key for the formation of galaxies, stars, and planets, is not understood and would require spatially resolved X-ray spectra providing detailed knowledge of the dynamics, temperature structure, and chemical composition of the hot atmospheric gas. Because the dense cooling gas and the hot phase need to be close to pressure equilibrium, information on the hot phase is crucial to understand the properties of the cooling gas observed at other wavelengths. 
\textbf{To measure the level of turbulence, which may play a key role in the formation of thermal instabilities, and to quantify the velocities and chemical make-up of both inflows and outflows close to the virial radii of galaxies, thus providing a final answer regarding feedback and the circulation of baryons in and out of $L^\star$ halos, several improvements with respect to the capabilities of {\it Athena} are essential.}

The measurement of gas velocities of around $100$ km/s, as well as the detection of very faint soft X-ray emission in the outskirts of galaxies will require a high spectral resolution of $E/\Delta E = 2000$ at 0.6 keV, a very large photon collecting area, and a low detector background. All these must be achieved while (at minimum) maintaining \textit{Athena}'s spatial resolution of $\sim5$ arcsec, which allows us to separate the emission of the inner CGM from that of the ISM and WHIM on scales of $\sim10$~kpc at z=0.1. A large FoV further offers the benefit of potentially probing several galaxy halos within a single observation, and increasing the studied sample size considerably. A future observatory with these capabilities will revolutionize our picture of the CGM and its link to galaxy evolution. 

\subsection{The role of galaxy groups in structure formation}
\label{sect:group}

\begin{mdframed}[backgroundcolor=blue!20] 
Groups of galaxies (defined as objects with $10^{13} < M_{500} < 10^{14}\: {\rm M}_\odot$) bridge the mass spectrum between galaxies and galaxy clusters. As such, they are critical systems for understanding the process of structure formation, the dynamical assembly of baryons in dark matter halos, and the complex physics that affects both the gas and the stellar components.
\end{mdframed}

Groups of galaxies are known to host a significant fraction of the number of galaxies in the Universe \citep[e.g.][]{Eke06}, and unlike galaxy clusters they form also in the filaments of the cosmic web rather than only in the nodes \citep[e.g.][]{Tempel14}.
Modern hydrodynamical simulations \citep[e.g.][]{LeBrun14,Planelles14,Truong18} show that the depletion of baryons in halos with $M_{500} < 2 \times 10^{14} M_{\odot}$ due to complex baryonic physics (e.g. cooling, galactic winds, AGN feedback) plays an important role in explaining the breakdown in self-similarity, wherein the $L_X-T_X$ relation shows a slope steeper than the expected value of 2 \citep[$\sim 3$,][]{Pratt09,Maughan12} at the cluster scale, further steepening at the group scale \citep[e.g.][]{Osmond04,Sun12}.

Understanding how the gas content of the outskirts of galaxy groups is affected by non-gravitational processes is needed to calibrate the baryonic effects on the matter power spectrum \citep[e.g.][]{hearin12,vanDaalen2015,vanDaalen2019} and is an important ingredient for deriving cosmological constraints from ESA's upcoming \textit{Euclid} mission.

In the coming years, \textit{eROSITA} \citep{Merloni2012} is expected to find over $10^5$ massive halos through their X-ray emission, a large fraction of them groups with masses below $10^{14} M_{\odot}$ \citep{borm14}, enabling cluster cosmology with an unprecedented statistical sample \citep[e.g.][]{pillepich18}. \textit{Athena} \citep{nandra13} will provide tighter constraints on the scaling relations and radial thermodynamic profiles of these objects \citep{ettori13,pointecouteau13}. 

However, studies with existing or approved missions, such as \textit{eROSITA} and \textit{Athena}, will be typically limited to the inner parts of galaxy groups. The mission envisaged here will combine both advantages, large grasp and high spectral resolution, thereby enabling a quantum leap in our study of the outskirts of galaxy groups.  In particular, the Cosmic Web Explorer will address better than ever before fundamental questions such as: \textbf{What is the baryon fraction out to $r_{200c}$ for the general population of galaxy groups as a function of radius and cosmic time? \textcolor{purple}{What are the roles of AGN feedback and non-thermal pressure in breaking self-similarity at the group scales?} 
How does this baryonic physics affect the matter power spectrum on non-linear scales? } 

\subsection{Filling the bridges, and emptying the voids, of the large-scale structure}
\label{sect:whim}

Most of the baryons in the local Universe are expected to be distributed along the filamentary structures that compose the backbones of the ``Cosmic Web''. Studies of UV-absorption lines with FUSE and HST-COS \citep{shull14} have probed the coldest fraction of these baryons. 
Hydrodynamical simulations \citep[e.g.][]{cen06} show instead that, much like for the CGM, the hotter phase is the dominant one (i.e. $T > 10^{5.5}$ K), and it can be detected preferentially through highly ionized C, N, O, Ne, and Fe ions in X-rays. 

Our understanding of how the growth of cosmic structures has developed requires that such baryons must have been processed by strong shocks at least once during their lifetime, where their infall kinetic energy has been mostly dissipated into thermal energy \citep[e.g.][]{1972A&A....20..189S,by08,2019SSRv..215...14B}. 
While there are hopes to detect at least the ``tip of the iceberg'' synchrotron signature of relativistic electrons accelerated by accretion shocks with the new generation of radio telescopes, the challenge of detecting such strong shocks (which are a pillar of our understanding of the WHIM picture) can only be tackled by finally imaging diffuse gas flows associated with $\sim 10^6-10^7$~K post-shock temperatures.

\begin{figure}[t]
\centering
\hbox{
\includegraphics[width=0.56\hsize]{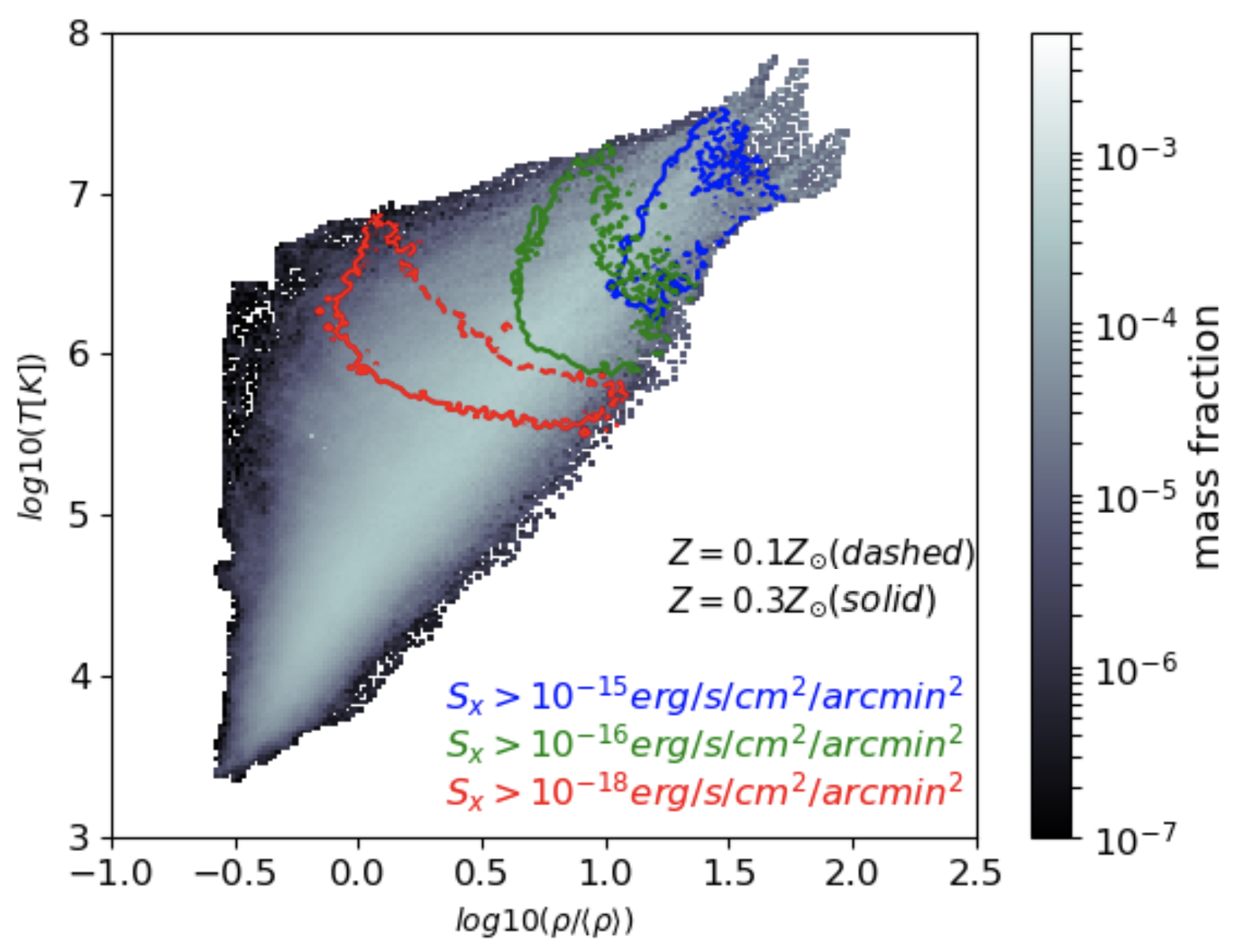}
\includegraphics[width=0.40\hsize]{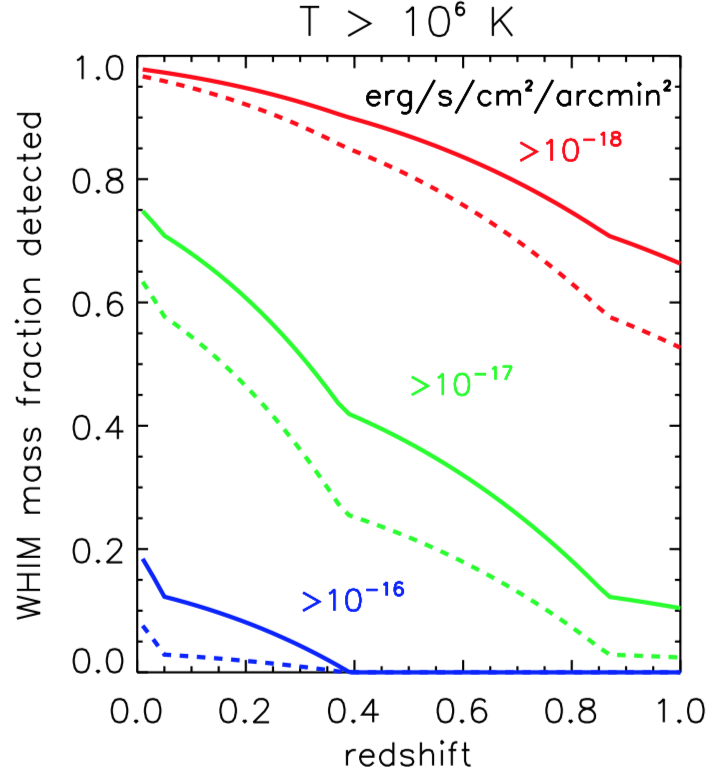}
}
\caption{\footnotesize{(Left) Phase diagram of the gas in a comoving 85 Mpc$^3$ volume, simulated using 1024$^3$ cells with the ENZO code \citep[][]{enzo14} and including the effect of magnetic fields, radiative cooling, star formation, supermassive black holes, and of the thermal feedback from star-forming regions and AGN as detailed in \citet{va17cqg}. Lines indicate the sensitivity required to observe it.
(Right) Distribution of the mass fraction of the gas at $T>10^6 K$ that is expected to be resolved, as a function of redshift, at different cuts in surface brightness in the 0.3--2~keV band, and for different metallicities. 
}}
\label{fig:whim}
\end{figure}

Away from higher-density regions that are being actively heated and stirred by complex stellar and AGN feedback, the truly diffuse, extended WHIM is a unique probe of structure formation processes and chemical enrichment history. Through both cosmic accretion and metal dispersion by feedback, the physical properties of the WHIM are a direct consequence 
of the interplay between the intergalactic medium (IGM), galaxies, and the action of gravity on much larger scales. 

\textbf{\textcolor{purple}{What is the structure of the Cosmic Web?}} 
To date, only a few reliable detections of the WHIM in absorption have been reported, using very long observations of bright, distant quasars \citep{nicastro2018}; the list is equally short for individual detections of LSS filaments in emission, which have typically been identified as they connect to the outskirts of massive galaxy clusters (see Section \ref{sect:clump}).
The \textit{Athena}/X-IFU will probe $\sim$200 of the strongest OVII absorbers associated with large-scale structure filaments, with typical properties expected to account for around 20\% of the WHIM baryon mass. In a few cases, the corresponding emission from the WHIM will also be detected.
Given \textit{Athena}'s spectral resolution in the soft X-ray band, such studies will be focused on the detection of the WHIM; its kinematics (and thus energetics) and, to a large part, its temperature and metallicity, will remain unexplored.
\textbf{We therefore need a mission with new capabilities that, in addition to spatially mapping the vast majority of the X-ray emitting cosmic web, will also routinely provide information about its detailed physical properties}.

\begin{figure*}[t]
\centering

\includegraphics[width=0.95\hsize]{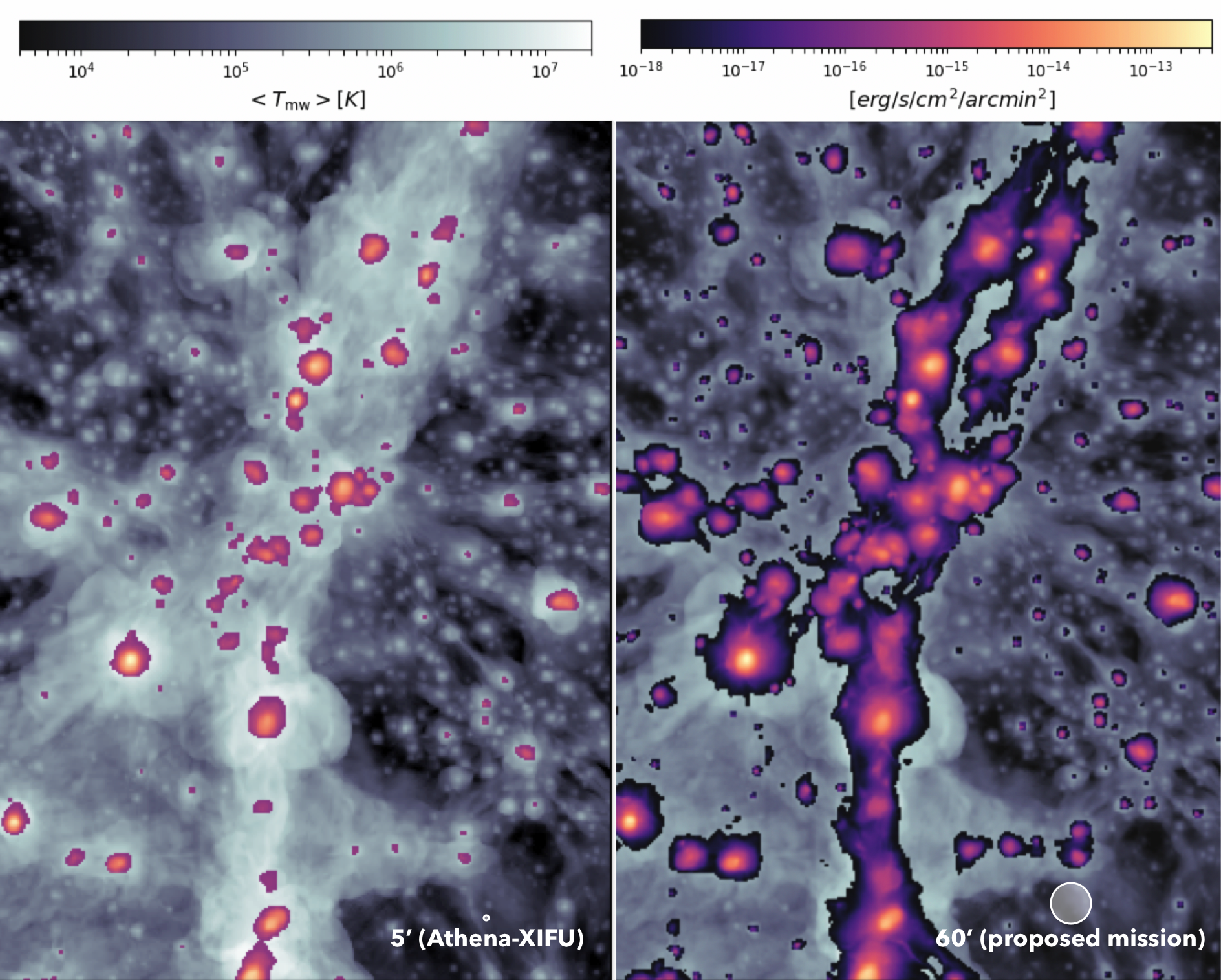}
\caption{\footnotesize{Projected mass-weighted gas temperature and detectable X-ray emission in the 0.3--2 keV energy band
at $z\approx0.05$ extracted from a cosmological hydrodynamical simulation \citep{vazza19}. Two different limits in surface brightness are used: (right) $\sim 1 \times 10^{-18}$ erg s$^{-1}$ cm$^{-2}$ arcmin$^{-2}$, which is the target for the proposed mission; (left) $\sim 5 \times 10^{-16}$ erg s$^{-1}$ cm$^{-2}$ arcmin$^{-2}$, which should be reachable by {\it Athena} for a similar investment in observing time, factoring in the difference in grasp and instrumental background.  
Each panel is $\sim 15^\circ \times 11^\circ$ across, about 10\% of the area that can be surveyed at this depth in 5 years with the proposed mission. 
The expected FoV for the {\it Athena}-XIFU (5 arcmin) and the Cosmic Web Explorer (1 deg) are shown in the lower right corner of both panels. }}
\label{fig:lss_maps}
\end{figure*}

Figure \ref{fig:whim} shows that a surface brightness sensitivity $S_X \sim 10^{-18}$ erg/s/cm$^2$/ arcmin$^2$ in the 0.3--2.0 keV band is sufficient to \textbf{detect \textit{all} of the diffuse baryons with a temperature $T>10^6$~K, and probe overdensities of $\rho/<\rho>\sim 1$}. For a WHIM component in collisional ionization equilibrium with $kT\sim0.2$~keV and a metallicity similar to the cluster outskirts (0.3 Solar), and modeling the foreground and background contributions as shown in Figure \ref{fig:emis_spectra}, we find that \textbf{an exposure time of 50~ks with the proposed mission would allow us to detect at least two emission lines (OVIII and one line in the OVII triplet) each with a significance greater than 5$\sigma$ per 1000 arcmin$^2$ extraction area. With a 5-year mission and given the observing efficiency in low-Earth orbit (LEO), 1600 square degrees can be surveyed at this depth}. Deeper observations can be used in the case that the metallicity and ionization state of the WHIM are different than these assumptions\footnote{Observations of 1~Ms can probe OVIII emission down to a limiting flux of $6\times10^{-12}$ photons/s/cm$^2$/arcmin$^2$.}. Figure \ref{fig:lss_maps} shows the dramatic improvement in our ability to map the WHIM using the proposed mission compared the \textit{Athena}/X-IFU, for a comparable investment of observing time. 

The expected signal is much weaker than the cosmic foregrounds and backgrounds. Priors on the spectral shape of the WHIM, together with detailed modeling of the foregrounds and noise, will be of paramount importance to extract the information we are searching for. This is a challenging task, but one where we can build upon improvements in spectral models of emission from our own Galaxy driven by \textit{XRISM} and \textit{Athena}, as well as the experience with data analysis techniques from cosmic microwave background (CMB) and gravitational wave science.

\begin{figure*}[tb]
\centering
\includegraphics[width=0.49\hsize]{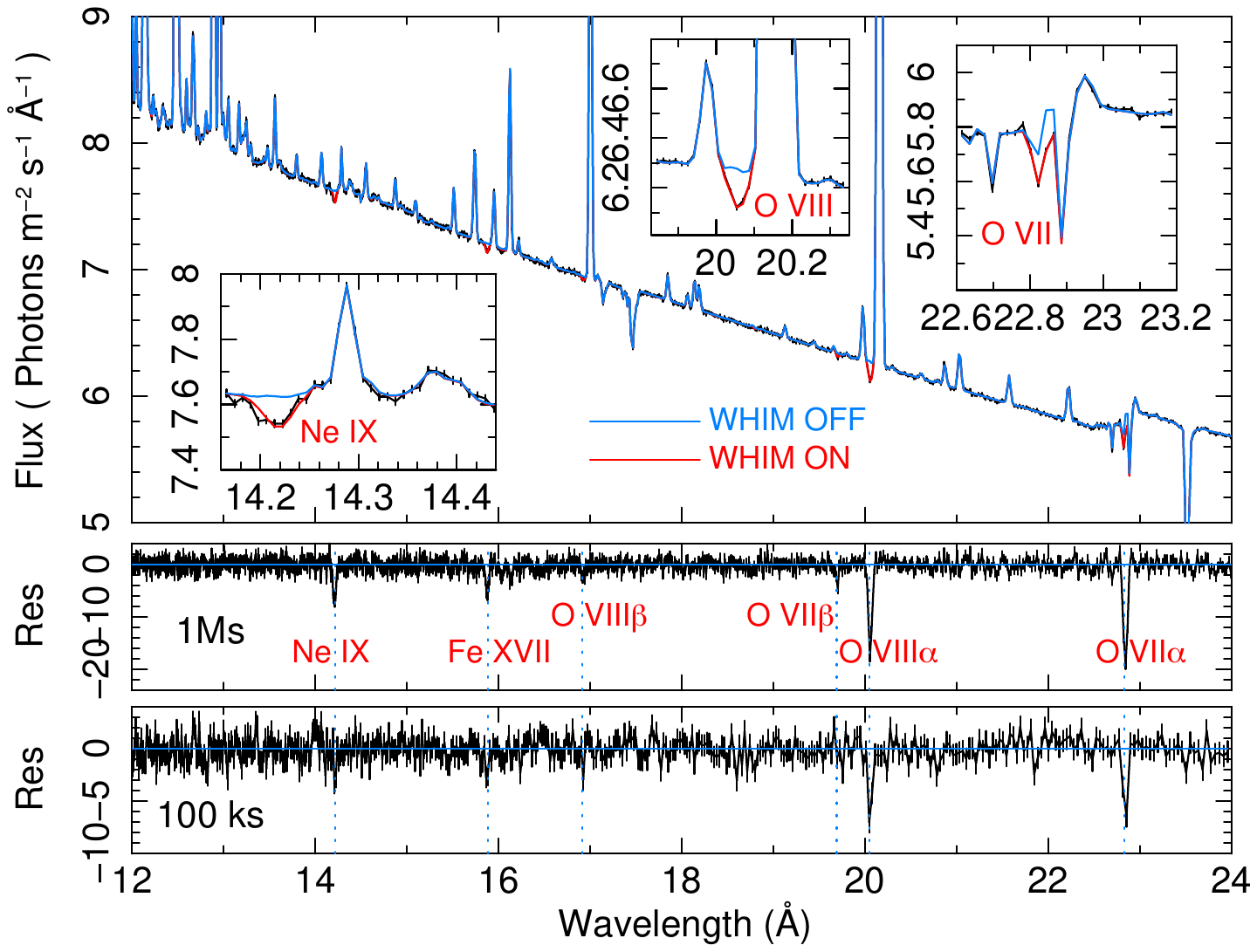}
\includegraphics[width=0.49\hsize]{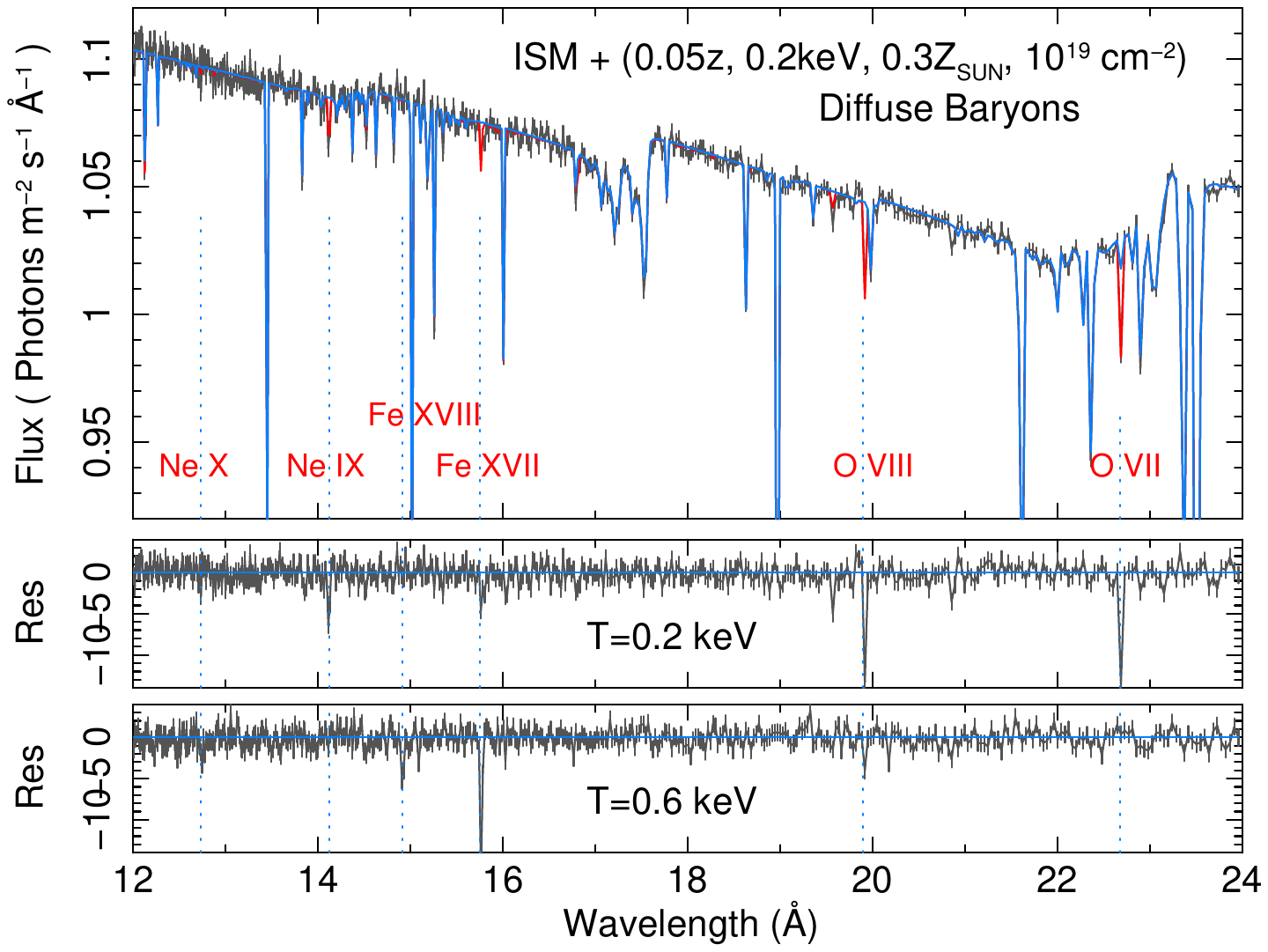}
\caption{\footnotesize{Simulations of faint, diffuse baryons that can be detected in absorption by the proposed mission. Left: an observation of a WHIM filament with kT=0.2 keV, Z=0.3 Solar, and $N_{\rm H}=10^{19}$ cm$^{-2}$, assumed to exist in projection in front of the cool core of A1795 at z=0.06, with a velocity difference of -1500 km/s from the cluster core. Both emission lines from the cool core and absorption lines from the WHIM can be studied with a 100~ks exposure (bottom panel residuals).
Right: more than 100 AGN with fluxes above $10^{-14}$ ergs/s/cm$^2$ are expected to be found within 1 square degree. Shown is a simulated, stacked spectrum of these AGN with an exposure of 1~Ms, with the same properties for the absorbing gas as in the left panel. A shorter 100~ks observation (not shown) would already result in a highly significant detection ($\Delta C_{\rm stat}\sim$50--80). 
For an absorber with a temperature of 0.6~keV (representative of the gas immediately inside the accretion shock of a massive galaxy cluster), the residuals for $N_{\rm H} = 5\times10^{19}$ cm$^{-2}$ absorbed by the same CXB are shown in the bottom right panel. Column densities below $1\times10^{19}$ cm$^{-2}$ at $kT=0.6$ keV can be reached in absorption against bright blazars.}}
\label{fig:abs_spectra}
\end{figure*}

Simultaneous emission and absorption spectroscopy will allow for direct, model-independent measurements of the gas density, length scale, ionization balance, excitation mechanism (or gas temperature), and element abundance. Absorption studies along independent lines of sight, numerous enough to minimize the impact of cosmic variance,
can set strong constraints on the WHIM properties. What makes the proposed mission unusually powerful compared to a grating spectrometer is that, in addition to mapping faint emission over large areas of the sky, \textbf{column densities as low as $N_{\rm H} \sim 10^{19}$ cm$^{-2}$ can be probed in absorption \textit{against the ubiquitous cosmic X-ray background}}, by stacking faint point sources detected in the field (see right panel of Fig.~\ref{fig:abs_spectra}). We therefore no longer have to rely on the serendipitous existence of a bright AGN within the region of interest to probe it. Moreover, only a non-dispersive X-ray spectrometer can use the diffuse ICM as a backlight for absorption studies. Clusters are among the brightest X-ray sources and reside at the nodes of the cosmic web; thus cluster sight lines are likely to pass through the densest regions of the WHIM. The proposed mission will study in great detail the absorption spectra from all WHIM filaments lying in projection in front of galaxy cluster cores (see left panel of Fig.~\ref{fig:abs_spectra}). 

Models based on hydrodynamical simulations \citep[see e.g.][]{borgani04} predict that WHIM filaments show a characteristic signal at angles of a few arcminutes that can be used to disentangle the WHIM from other components of the unresolved X-ray background \citep[see][]{roncarelli06a,hickox07}, with uncertainties related to its metal composition \citep[][]{ursino10,2011ApJ...731...11C,2012MNRAS.424.1012R}, and an associated high cosmic variance \citep[see e.g.][]{ursino14}.
This emission can be resolved as a signal in the angular auto-correlation function (or its analogue in the Fourier space, the power spectrum) on arcminute scales, although the peak in the signal will move to smaller angles for contributors at higher redshifts. Once the rich structure of the cosmic web is also resolved in redshift, higher order statistics, such as the 3-point correlation function and tools from graph theory \citep[e.g.][]{naidoo19}, can be applied for robust cosmological parameter inference using the densest structures.
Cross-correlations with other plausible signposts of the cosmic filaments, such as galaxy luminosity density \citep[see, e.g.,][]{nevalainen15} and diffuse inverse Compton scattering of the CMB photons on the WHIM in mm bands \citep[see, e.g.,][]{tanimura19}
will enhance the signal associated with the WHIM emission.
\cite{ursino14} show that the cross-correlation of the WHIM signal in the soft (e.g. 0.4--0.6 keV) X-ray and SZ maps is dominated by the gas within $z < 1.5$ and peaks at scales of a few arcmin (multipole $l \sim 10000$).
Combining radio and X-ray observations will also allow us to detect the cosmic web illuminated by structure formation shocks \citep{vazza19}.
Gas ``bridges'' tracing the leading merger axis between clusters in an early merging stage are boosted both in radio and X-ray emission, compared to the more typical conditions found in cluster outskirts. These boosted emission regions should be already visible with existing radio instruments \citep[e.g. LOFAR; see][]{govoni19}, and can be studied in X-rays with \textit{Athena}. In the future, the increased sensitivity of the SKA, together with the improved capabilities of the proposed Cosmic Web Explorer, can push this type of joint exploration towards fainter and more representative WHIM filaments.

A role complementary to the filamentary structure of the cosmic web is played by the cosmic voids. Devoid of matter by definition, they are dark energy-dominated objects, and are particularly sensitive to neutrinos (and all diffuse components) since the mass fraction of neutrinos with respect to CDM is higher in voids than in high density regions. The evolution of voids is ruled by the joint action of gravitational attraction, that empties voids by pushing material towards their boundaries, and the expansion of the Universe, that also enlarges voids by diluting the space between galaxies.
For these reasons \citep[see, e.g.,][]{pisani19}, observables such as number, size, shape, distribution and clustering of cosmic voids are powerful probes of the properties of the dark energy and neutrinos.

\begin{mdframed}[backgroundcolor=blue!20] 
Tomographic observations of the soft X-ray emission from contiguous patches of the sky will allow the reconstruction, both directly and, at higher statistical significance, with auto-correlation function techniques, of the distribution of the WHIM at $T>10^6$~K in different redshift bins (see Fig.~\ref{fig:whim}). 
The combination of the signal on the sky at various redshifts will permit a first 3D mapping of the filaments (and relative voids) containing most of the elusive fraction of the missing baryonic mass.
\end{mdframed}

\begin{table}[!ht]
\centering
\caption {Driving scientific goals behind each of the mission requirements. Implicitly, this also shows which science cases would be most affected if any of the proposed capabilities were reduced to achieve a smaller mission profile.}
\vspace{0.3 cm}
\small
\begin{tabular}{|>{\raggedright}p{2.cm}|>{\raggedright}p{2.cm}|p{7cm}|}
\hline
\textbf{Parameter} & \textbf{Value} & \textbf{Science driver}  \\ \hline
 Field of view                   & 1 deg$^2$  & Mapping the connection between galaxy clusters and the large-scale structure; surveying cosmic web filaments; cross-correlation with galaxy distribution, weak lensing maps, SZ surveys over a wide area. \\
  & & {\bf Athena XIFU is designed with a field of view diameter of 5 arcmin, allowing detailed study of the gas kinematics in clusters only along pre-selected azimuthal sectors.} \\
\hline
 Angular resolution (90\% enclosed energy fraction)     & $\approx 5^{\prime\prime}$  &  Resolving structure in the circum-galactic medium; separating galaxy halos from the truly diffuse LSS filaments; removing contamination from point sources to a sufficient depth to enable the study of very faint diffuse emission. \\
  & & {\bf Athena's PSF is expected to be comparable.} \\
\hline
 On-axis effective area          & $\approx$ 10 m$^2$ @ 1 keV &  Detection and mapping of cosmic web filaments at low overdensities. \\ 
   & & {\bf Athena is designed with an effective area a factor of $\sim$7 lower at 1 keV.} \\
   \hline
 Energy band                     & 0.1--3 keV  & Studies of the chemical composition from C to Ni, up to and including K-shell lines of Si. \\ 
  & & {\bf Athena is expected to be sensitive to energies up to $\sim$10 keV that are not needed for the scientific case of the Cosmic Web.} \\
  \hline
 On-axis sensitivity   & $10^{-18}$ erg cm$^{-2}$ s$^{-1}$ arcmin$^{-2}$ & Detecting and characterizing the cosmic large scale structure in emission at $T>10^6$K.\\
   & & {\bf The combination of effective area and background reproducibility would allow Athena to reach a surface brightness limit of a few times $10^{-16}$ erg/s/cm$^2$/arcmin$^2$ in the 0.5--2 keV band (see for a comparison Fig.~\ref{fig:lss_maps}).} \\
\hline
 Spectral resolution             & E/$\Delta$E = 2000 at 0.6 keV & Measuring kinematics of the CGM.  Efficiently detecting absorption lines for low column densities. Separating the expected signal from the Milky Way halo emission. \\
    & & {\bf Athena X-IFU is designed to reach a resolving power of a few hundreds in the soft band.} \\
\hline
 Detector background            & $\leq 1.5\times10^{-4}$ cts s$^{-1}$ keV$^{-1}$ arcmin$^{-2}$
 & Constraining the electron number density in faint, diffuse plasma from the bremsstrahlung continuum. Efficiently detecting faint emission lines against the continuum level.\\
    & & {\bf The Athena X-IFU design has a nominal internal background of
    $\sim 5\times10^{-3}$ cts s$^{-1}$ cm$^{-2}$ keV$^{-1}$, corresponding to $\sim 6.1 \times 10^{-4}$ cts s$^{-1}$ keV$^{-1}$ arcmin$^{-2}$ for a 12m focal length. } \\
\hline
 Mission duration               & 5 years & To cover $\sim$1600 square degrees with deep observations ($\sim50$~ks), given observing efficiency in LEO. \\
    & & {\bf Athena is designed for a nominal duration of 4 years.} \\
\hline
\end{tabular}
\label{tab:table_req}
\end{table}

\section{Mission concept}
\label{sec:concept}

\begin{figure}
\centering
\includegraphics[width=0.8\textwidth]{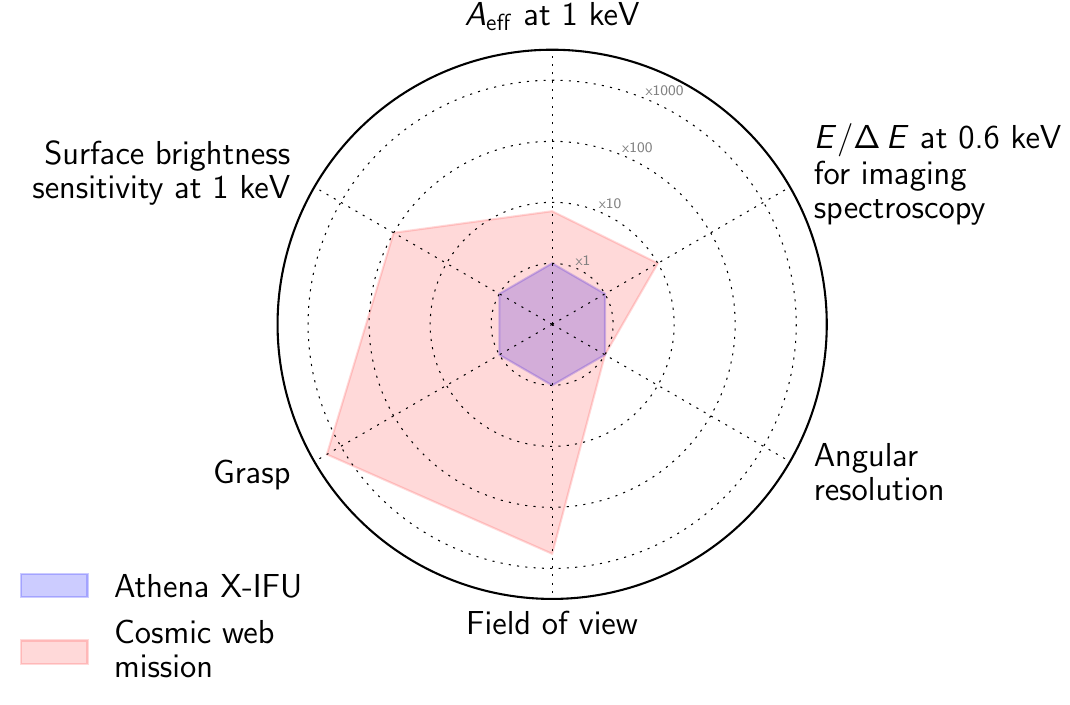}
\caption{\footnotesize Enhancements in the capabilities of the proposed Cosmic Web Explorer with respect to the {\it Athena} X-IFU. Each dotted circle represents an order of magnitude increase.} \label{fig:radarplot}
\end{figure}

\color{black}
\subsection{Overview}

The science questions described above can be answered with a mission providing the necessary combination of throughput, grasp, angular resolution, spectral resolution, and low and stable detector background. The basic mission concept is thus an L-class, large effective area X-ray telescope with a large field of view, focusing on an X-ray Integral Field Unit, and placed in Low-Earth orbit. Table~\ref{tab:table_req} lists the driving scientific goals and associated mission requirements. The envisioned improvements compared to the \textit{Athena} X-IFU are shown in Figure \ref{fig:radarplot}.

A small fraction of the science theme proposed here, namely probing the CGM, WHIM, and cluster outskirts only along selected sightlines towards bright background AGN, can be covered by an ESA Medium-Class mission based on dispersive technology spectrometers \citep[see][]{nicastro2050wp}. 

\subsection{Mirror payload}


\subsubsection{Segmented glass foils}
Glass can be shaped to high accuracy and low roughness. Glass foils have a low density (2.3~g/cm$^3$) and can be manufactured down to a 0.2~mm thickness, based on the experience matured with NuSTAR \citep{Harrison13}. This enables the minimization of obstructions and of the mirror module mass. 

The initial design based on segmented glass foils foresees the determination of the mirror module diameters and the focal length, aiming at reaching an effective area close to 10~m$^2$. To enhance the effective area at low energies, it is important to mitigate the photoelectric absorption by using a coating of amorphous carbon \citep{Pareschi04, Cotroneo07}. 
Here, we suggest as a guideline two different possible solutions, based on segmented glass shells and an Au+C coating. Alternative coatings such as Ir+C or Ir+SiC, or the use of various coatings for different mirror shells, can be optimized as the design is developed in more detail. Also, the current design assumes a Wolter-I mirror profile. To minimize off-axis degradation of the point spread function (PSF), Wolter-Schwarzschild or polynomial designs \citep{Conconi2001} should be investigated at a more advanced stage of the mission definition.

\paragraph{Solution 1: single mirror module, 20~m focal length}
The first possible option is represented by 310 coaxial shells, with focal lengths of 20~m and diameters ranging from 4.3~m down to 1.492~m. There is no relevant benefit in extending the series beyond these limits. The shells are kept as close as possible, in order to minimize the dead areas between shells, but at the same time ensuring a FoV (50\% vignetting function) of 1~deg in diameter. The mirror thickness is fixed at the value of 0.4~mm, and the total mirror mass is nearly 1000~kg. 

\begin{figure}
\centering
\includegraphics[width=0.8\textwidth]{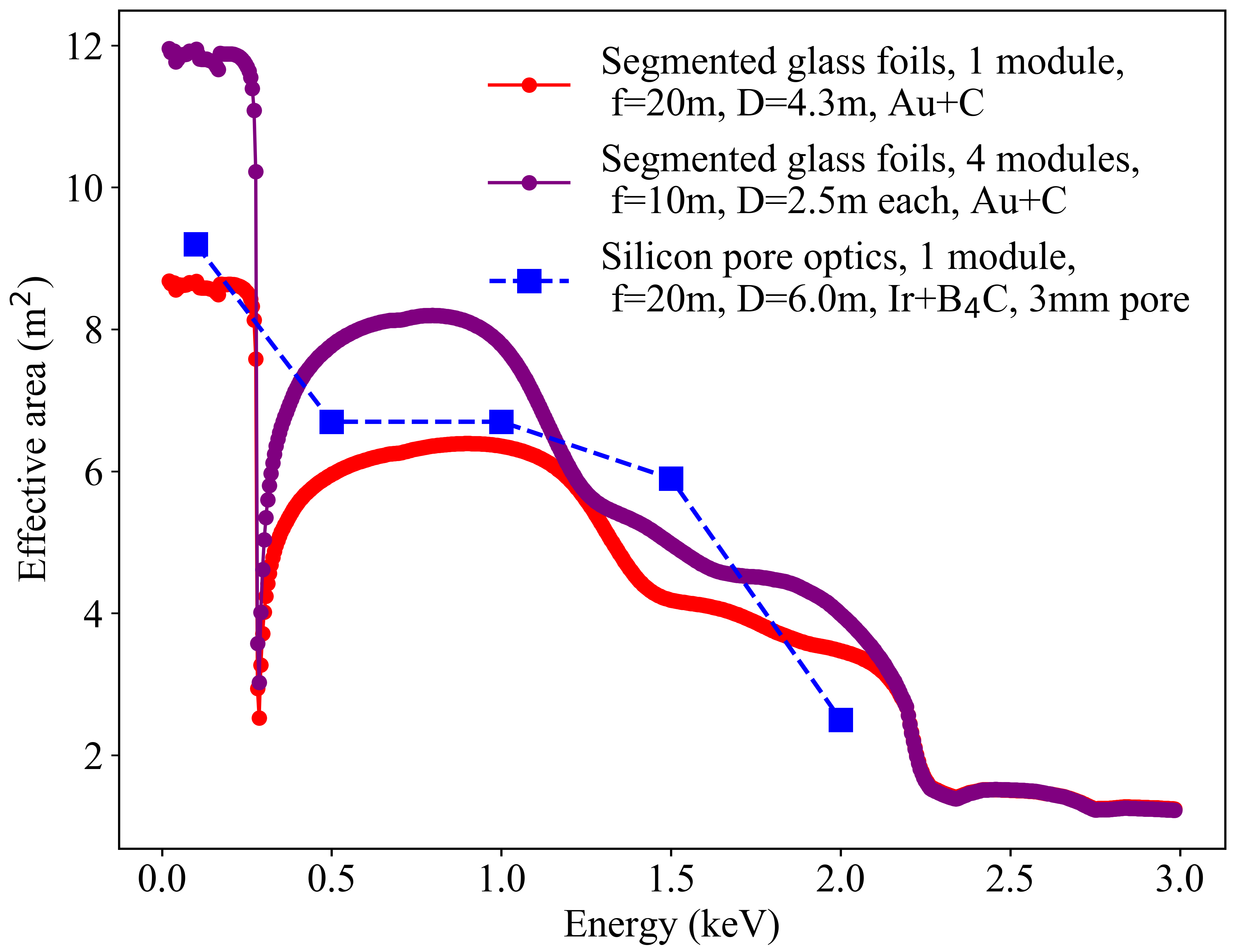}
\caption{\footnotesize Effective area expectations for several designs using segmented glass foils or SPO.} \label{fig:eff_area1}
\end{figure}

 Figure~\ref{fig:eff_area1} shows the expected effective area with an Au+C coating, 150~nm Au, and 30~nm C (red dashed line, assuming a further 10\% of obstruction due to supporting structures). We can reach $\sim$6 m$^2$ around the line energies of OVIII and Fe XVII. A coating with bare gold would lead to some gain around 2~keV, but at the expense of an effective area loss near 1~keV.

\paragraph{Solution 2: four mirror modules, 10~m focal length}
In order to shorten the focal length and reduce the detector background, while possibly increasing the effective area, we can divide the optical module into four identical sub-modules, each of them imaging onto a separate detector. If a fairing diameter of 6~m can be envisaged (commercial launch providers such as Blue Origin and SpaceX plan to offer 7~m and 9~m fairings in the next few years), four identical mirror modules with a 2485~mm diameter can be accommodated. We consider a 10~m focal length and populate each module with decreasing diameters, keeping appropriate spacing between shells to preserve the geometric FoV. In this way, we can extend the series down to a 766~mm radius through 185 mirror shells. The total mirror mass now is $\sim$1350~kg.

Figure~\ref{fig:eff_area1} shows the comparison between the 1-module and the 4-module design. The 4-module concept enables higher effective area performances, with a considerable gain near 1~keV. An analytic computation \citep{Spiga11} of the off-axis area at 0.2~keV and 1~keV showed that the FoV is very close to 1~deg also with this design. 

\subsubsection{Silicon pore optics} 

Silicon pore optics (SPO) are a new technology for the construction of segmented X-ray optics of large diameter which is the baseline technology for ESA's $Athena$ telescope. 
SPO are made from ribbed silicon plates obtained from industry-standard silicon wafers
that are shaped with the help of a high quality mandrel to obtain X-ray optical elements
called stacks. These stacks are then used as the primary or secondary mirror in an approximation of the Wolter-I geometry. 
The requirements of the Cosmic Web Explorer mission are demanding, and in order to meet the specifications SPO technology will need to be further developed. Both a large outer radius of the mirror and wider pores are needed to achieve the large field of view. 

We have considered a range of mirror designs with focal lengths of 10, 12, and 20 m, assuming a coating of 10nm Ir and 4nm B$_4$C, in line with the development activities of \textit{Athena}. The 20 m options are the most attractive from the point of view of the effective area and the required plate lengths. In Figure~\ref{fig:eff_area1} we show the effective area that can be obtained for a mirror design with a 20 m focal length, 3 mm pore width (compared to 2.3 mm for $Athena$), 20 mm plate length, and an outer radius of 3 m. For this design, the SPO mechanical vignetting reaches 50\% at an off-axis angle of 45 arcmin.

Shorter telescope focal lengths would require, in addition to the developments mentioned above, the use of plates that are just 10 mm long, i.e. half of the shortest plate available so far. A possible alternative to this, which is currently being investigated, is the creation of stacks where the plates can be spaced by more than a silicon wafer thickness, and therefore can be made longer. Pursuing this technological progress is important for realizing the science goals of this mission, if it were selected.

\subsection{Detector payload}

An integral field unit (IFU) provides the optimal combination of resolving power ($\Delta E/E \sim 2000$) and spatial resolution imaging (PSF $\leq 5^{\prime\prime}$) for the science theme. 
The resolving power requirement can be met by the use of a cryogenic X-ray microcalorimeter array. A number of technologies are under active development for the construction of such arrays, including metallic magnetic calorimeters \citep[MMC, e.g.][]{mmc18}, metal insulator sensors \citep[MIS, e.g.][]{sau16,sau18}, and transition edge sensors \citep[TES: for reviews, ][]{IrwinHilton, TES_review15}. 

\begin{figure*}[t]
\centering
\includegraphics[width=\textwidth]{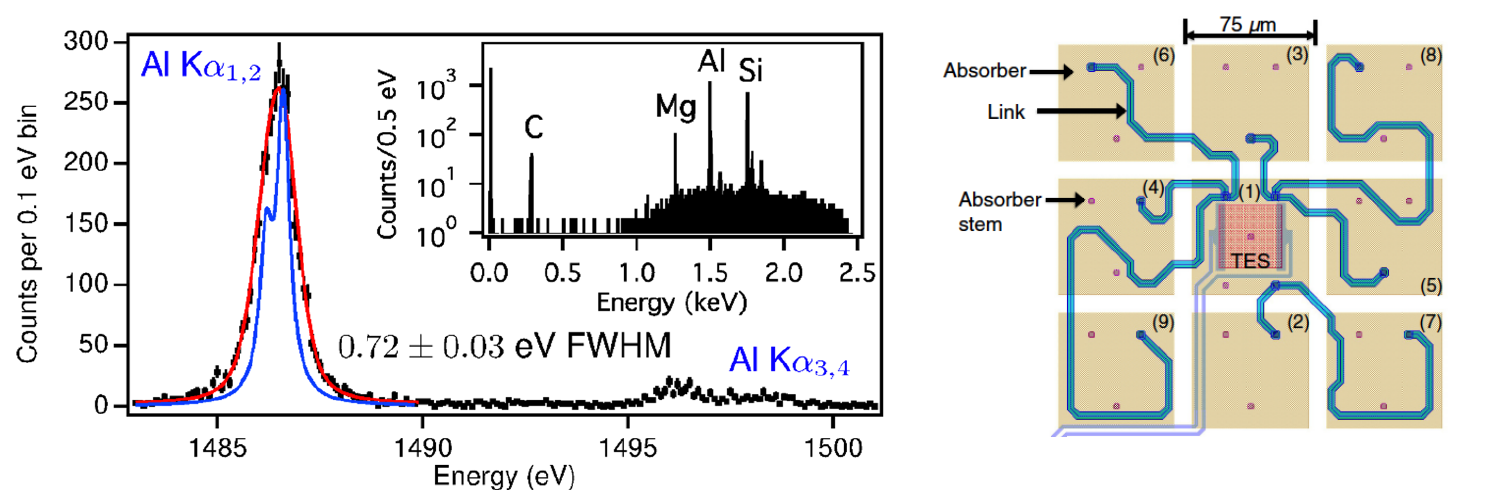}
\caption{\footnotesize (Left:) 
Example measured spectrum of the Al K$\alpha$ complex.
The blue curve is the intrinsic line shape, and the red curve is the best fit to the spectrum, demonstrating that a resolution of 0.7~eV is already achieved in the laboratory \citep[][]{lowE15}.
(Right:) Schematic layout of a 9-pixel $hydra$~\citep[][]{hydra19}. Multi-absorbers are connected to one TES with different thermal conductance (blue line), effectively reducing the number of TESs that need to be read out.
}\label{fig:tech}
\end{figure*}

As an example, the spectral performance of TES calorimeters can be described as
\begin{equation}
\Delta E\sim2.35\sqrt{\frac{k_BT^2C}{\alpha}}, ~\mathrm{or}~ \Delta E\sim2.35\sqrt{{k_BTE_{\mathrm{Max}}}},
\end{equation}
where $k_B$, $T$, $C$ and $\alpha$ are the Boltzmann constant, detector temperature, heat capacity, thermistor sensitivity, respectively, and the saturation energy can be written as $E_{\mathrm{Max}}=CT/\alpha$. 
Based on the scaling from the state-of-art performance of TES calorimeters, 0.5-0.7 eV $@$ 1.5 keV in laboratory conditions \citep[][and Figure \ref{fig:tech}, left]{lowE15}, it is feasible to achieve $E/\Delta E\sim2000$ assuming $E_{\mathrm{Max}}\sim0.6$ keV and $T\sim60-70$ mK. 

\begin{figure*}[t]
\begin{center}
\includegraphics[width=\hsize]{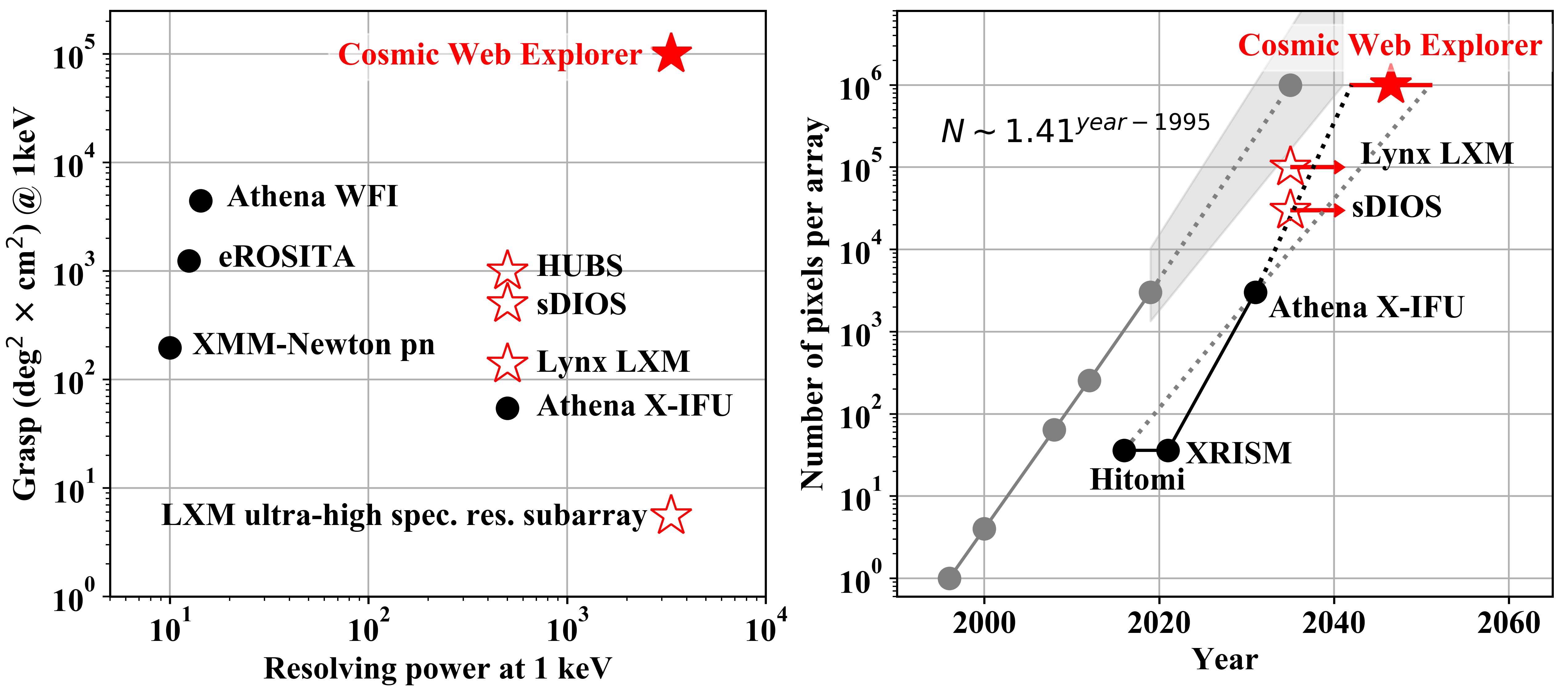}
\end{center}
\vspace{-0.5cm}
\caption{\footnotesize (Left)  Grasp versus spectral resolution at 1~keV for selected operational/accepted (black circles) X-ray observatories and proposed mission concepts (red stars). 
(Right) Number of X-ray/gamma-ray calorimeters per instrument in laboratory environments (gray circles) vs.\ space (black circles for past or upcoming X-ray micro-calorimeter missions at their respective launch year, red stars for a subset of mission concepts illustrating the expected future progress; horizontal red lines/arrows represent launch date uncertainties). The shaded gray area shows the extrapolation of a power-law function ($N\sim\rm {1.41}^{year-1995}$) fit to the laboratory data, with conservative uncertainties.}
\label{fig:grasp}
\end{figure*}

A large number of pixels would be needed to cover the wide FoV. Assuming a similar detector pixel size (pitch) as that of the {\it Athena} X-IFU (275 $\mu$m$^2$: $\sim$5 arcsec$^2$), it would be necessary to have 3000$\times$60$^2$/5$^2\sim$4$\times10^5$ pixels to cover a FoV with a diameter of one degree; ideally, this number could be further increased to avoid under-sampling the telescope's PSF. Such a quasi-mega-pixel array would represent a thousandfold increase in resolution elements compared to {\it Athena}, and even an order of magnitude more than that envisaged for
the {\it Lynx} X-ray Surveyor, a high-energy flagship mission concept being considered for the NASA 2020 Astrophysics Decadal Survey (see Fig. \ref{fig:grasp}, right). Several research and development (R\&D) components will need to be addressed to make such a megapixel calorimeter array a reality. Some examples are shown below.

\paragraph{New multiplexing readout technology}
Due to the strict limit of available electrical and cooling power in space, the multiplexing readout technology is one of the key R\&D items required for the proposed project. 
One of the most promising multiplexing technologies currently under development is the \textit{Microwave SQUID multiplexer} \citep[$\mu$-wave MUX:][]{uwave04, uwave08}. The $\mu$-wave MUX consists of a number of superconducting resonators in the GHz range, each employing a unique resonance frequency, terminated by an inductance magnetically coupled to dissipationless rf-SQUID. This technology allows the readout of a number of TES pixels that is an order of magnitude larger than conventional multiplexing approaches.

\paragraph{Detector fabrication technology}
Even using $\mu$-wave MUX technology, it can still be challenging to read out 10$^6$ TES pixels. In this regard, multi-absorber TES (Fig.~\ref{fig:tech} right), in a so-called $hydra$ configuration, will increase the technological feasibility. Absorbers have different thermal conductance to TES. Therefore, even for the same incident photon energy, the response of each absorber is different. By using this difference of the response, the arrival information (i.e. which absorber recorded a photon hit) can be extracted. The $hydra$ configuration effectively reduces the number of TESs. 

Since the original concept of using a thermal detector as an X-ray spectrometer~\citep[][]{mmm84} proposed about 35 years ago, we are now halfway to our goal leading up to the year 2050. Considering the progress made so far, we are confident that, building upon the advancements driven by the \textit{Athena}/X-IFU instrument design,
all technological developments are feasible with continuous effort.   

\paragraph{Optical/thermal blocking filters}
Last but not least, here we note the necessity of the future development of new thermal/optical filters, which are a critical component to avoid the performance degradation due to radiation shot noise. Because our targets primarily emit soft X-rays, the loss of the effective area due to the filter transmission is of importance. For instance, the thermal/optical filters for the $Athena$ X-IFU \citep{barbera18} will provide a transmission of $\sim$0.5 around 600 eV. This means the total effective area at 600 eV is reduced by one half of that provided by novel optics / enhanced mirror technology. Therefore, for our purposes, further development and optimisation of the optical/thermal filters compared to the current generation are required.

\newpage 

\section{Synergies in the context of astronomy into the 2050s}
\label{sec:synergies}

\begin{mdframed}[backgroundcolor=blue!20] 
For the science questions described above, there is a strong synergy between the proposed mission and many other facilities, both ground- and space-based, that will be dedicated to exploring galaxy overdensities, dark matter, and the gas phase as tracers of the Cosmic Web structure, and span the entire spectral range from X-ray to radio. 
\end{mdframed}
\color{black} 

~\\[-3em]

\paragraph{The hot and warm phase of the Universe} 
Beyond \textit{eROSITA}, \textit{XRISM}, and \textit{Athena}, several mission concepts are under study.
The Hot Universe Baryon Surveyor (HUBS\footnote{ http://hubs.phys.tsinghua.edu.cn/en/}) and the Diffuse Intergalactic Oxygen Surveyor \citep[][Super DIOS]{SuperDIOS2018}, with a much lower sensitivity and grasp and lower resolving power compared to that of the proposed mission (see Fig. \ref{fig:grasp}, left), will pave the way for a deep exploration of the warm gas in the Cosmic Web over large areas of the sky.
The Advanced X-ray Imaging Satellite \citep[][AXIS]{AXIS_SPIE} and the {\it Lynx} X-ray Surveyor \citep{Gaskin2017} are expected to provide a complementary, exquisite spatial resolution (down to $0.5$ arcsec), allowing us to map substructure in the gas with very high detail. 

In the coming two decades, projects such as CMB-S4 \citep{aba19}, the \textit{Simons} Observatory \citep{sim19}, and \textit{LiteBIRD} \citep{haz19} are expected to build up our knowledge of massive halos from the SZ effect and push the investigation of the millimetre CMB sky a step further. Beyond the 2030s, a high resolution, high sensitivity large millimetre telescope such as the AtLAST \citep{AtLast19} and CMB-HD \citep{CMB-HD19} concepts would bring opportunities that complement spectroscopic X-ray measurements. These features, generalised to a survey over a large area of the sky, would lead to a wealth of constraints on the thermodynamical properties (via the thermal and relativistic SZ effects), the peculiar motions (via the kinetic SZ effect) or the mass (via CMB lensing) of large scale structures. Such a mission concept is being proposed in the framework of Voyage 2050 \citep{basu2050wp}.
The strong connection between X-ray and SZ for the physical characterisation of hot plasmas is already well established \citep{Bulbul2019,mro19}, meaning that these future SZ observatories offer crucial synergies with the mission proposed here. Both observables will provide a complementary view of the thermodynamics and kinematics of warm-hot plasmas in the CGM, ICM, and WHIM out to the epoch of massive halo formation ($z\sim 2-3$).

The Large UV Optical Infrared Surveyor (LUVOIR) would provide another excellent complement to our proposed mission concept. LUVOIR's UV spectroscopy capabilities are uniquely poised to bring about a revolution in our understanding of gas flows, enrichment, and ultimately galaxy evolution on a wide range of cosmic scales \citep{LUVOIR}. 
The combination of the cool-hot UV and warm-hot X-ray emission lines will deliver a comprehensive picture of the complex physical phases of the baryons filling the LSS. This will allow us to probe the composition and physical processes that define gaseous halos over the mass spectrum, delivering a transformative understanding of galaxy evolution, galaxy cluster physics, and gas within the Cosmic Web. 

\paragraph{The light and the dark phase of the Universe} 
In the next decades, several optical and IR telescopes will survey the distribution of galaxies and quasars over very large patches of the sky: 
{\it DESI}\footnote{https://www.desi.lbl.gov} has completed, and released to the public in January 2021, the Legacy Imaging Surveys from which 35 million galaxies and 2.4 million quasars will be selected as targets for a 5-year-long spectroscopic follow-up set to commence after a re-commissioning due to the pandemic has been finalized; 
{\it Euclid}\footnote{https://www.euclid-ec.org/} will be launched in 2022; the $Rubin$ Observatory\footnote{https://www.lsst.org/} (previously known as LSST) will start a 10-year campaign at full regime in 2023, when {\it SPHEREX}\footnote{http://spherex.caltech.edu/index.html} will begin the first all-sky spectral survey between $0.75$ and $5\ \mu$m. By 2050, all of these facilities will produce catalogues of tens of millions of galaxies and, through their overdensities and measured gravitational shear, will map the distribution of dark matter and stars in the LSS over most of the sky and out to the epoch of reionization, locating the regions where baryons in the hotter phase are expected to reside. 

\paragraph{The neutral and non-thermal phase of the Universe} 
On the radio side, the leading long-term future facility is the Square Kilometer Array (SKA). Its forcasted Phase 2 improvements in sensitivity below $\leq 200$ MHz beyond 2030  
will greatly benefit the quest for low surface brightness radio structures in the cosmic web. 
According to recent numerical simulations, an improvement in sensitivity of a factor $\sim 3$ should result in a systematic detection of radio emission associated with accretion shocks in the cluster outskirts, as well as in imaging the radio-brightest regions of the larger cosmic web \citep[][]{vazza19}. 
Rare large scale filaments could be detected through the 21cm neutral hydrogen emission line  \citep[][]{2017PASJ...69...73H}.
The spectroscopic HI galaxy survey is also expected to detect millions of halos, competitive with surveys like $Euclid$. 
A reliable catalogue of targets expected to be filled with the most rarefied plasmas can be built upon these observations, enabling the physical properties of those plasmas to be revealed at last by further studies, e.g., in X-rays using future missions such as our proposed Cosmic Web Explorer.

\section{Conclusion}
\begin{mdframed}[backgroundcolor=blue!20] 
We propose a mission concept that provides the necessary combination of X-ray capabilities 
to answer the most fundamental questions on the distribution, energetic budget, and metal enrichment of the hot diffuse matter permeating the Universe's large-scale structure:
\begin{itemize}
    \item What are the properties of the soft X-ray emitting halos around $L^\star$ galaxies where most of the stars and metals in the Universe were formed? (see Sect.~\ref{sect:cgm})
    \item How does the dynamical assembly of baryons in dark matter halos occur over the entire mass spectrum, from $L^\star$ galaxies to galaxy groups to the more massive galaxy clusters? (see Sect.~\ref{sect:cluster}, \ref{sect:cgm}, \ref{sect:group})
    \item What are the complex physical processes affecting both the gas and the stellar components during the accretion of these baryons from the field into virialized structures like galaxies, galaxy groups and clusters? (see Sect.~\ref{sect:cluster}, \ref{sect:cgm}, \ref{sect:group})
    \item What is the role of the poorly understood non-equilibrium physics, which is responsible for the formation and evolution of cosmic structures, in the regions near, and beyond, the virial radii of galaxy groups and clusters?
    How are they fundamentally different from the physics in the cores of clusters that has been the focus of X-ray cluster science over the past decades? 
    (see Sect.~\ref{sect:cluster})
    \item What are the chemical and thermodynamical properties of the Warm-Hot Intergalactic Medium? How, and where, is it distributed in the local Universe?  (see Sect.~\ref{sect:whim})
\end{itemize}
\smallskip
\noindent To reach these goals, starting from the heritage of the ESA L-class mission \textit{Athena}, we need (see Table~\ref{tab:table_req} and Sect. \ref{sec:concept}):
\begin{itemize}
    \item a large effective area ($\sim 10$ m$^2$ at 1~keV);
    \item an X-ray Integral Field Unit array with an improved spectral resolution ($E/\Delta E = 2000$) in the soft X-ray band (0.1--3.0~keV);
    \item high sensitivity ($10^{-18}\,{\rm erg\,cm^{-2}\,s^{-1}\,arcmin^{-2}}$), thanks to an unprecedented control on the systematics of the X-ray background that will benefit from a Low-Earth orbit;
    \item a spatial resolution of 5~arcsec, similar to one achieved by \textit{Athena}, but over a large ($\sim 1$ deg$^2$) FoV. 
\end{itemize}
\noindent The combination in a single instrument of all these capabilities will permit a giant leap forward in the complete and exhaustive understanding of the physical processes that shape and define the Cosmic Web, providing the perfect synergetic complement in X-rays to facilities that will be available in other wavebands by 2050, in a joint and harmonious effort to know our Universe in depth.

\end{mdframed}
\color{black} 

\begin{acknowledgements}
We thank F. Nicastro, J.S. Kaastra, G. M. Voit, M. Donahue, J. Green, W. Cui, N. Hatch, D. Fielding, J. Sayers, J. P. Breuer, L. di Mascolo, F. Mernier and J. Croston, in no particular order, for fruitful discussions and support towards preparing this manuscript. 
A.S. gratefully acknowledges support by the Women In Science Excel (WISE) programme of the Netherlands Organisation for Scientific Research (NWO). 
S.E., M.R. and F.G. acknowledge financial contribution from the contracts ASI-INAF Athena 2015-046-R.0, ASI-INAF Athena 2019-27-HH.0,
``Attivit\`a di Studio per la comunit\`a scientifica di Astrofisica delle Alte Energie e Fisica Astroparticellare''
(Accordo Attuativo ASI-INAF n. 2017-14-H.0),
and from INAF ``Call per interventi aggiuntivi a sostegno della ricerca di main stream di INAF''.
D.N. acknowledges Yale University for granting a triennial leave and the Max-Planck-Institut f\"ur Astrophysik for hospitality. 
GWP acknowledges support from the French space agency, CNES. 
B.M. acknowledges support from the UK STFC under grants ST/R00109X/1, ST/R000794/1, and ST/T000295/1. F.V. acknowledges financial support from the ERC  Starting Grant ``MAGCOW'', no. 714196, the usage of Piz Daint supercomputer at CSCS-ETHZ (Lugano, Switzerland) under project s805, and the usage of online storage tools kindly provided by the INAF Astronomical Archive (IA2) initiative (\url{http://www.ia2.inaf.it}). 
VB acknowledges support by the Deutsche Forschungsgemeinschaft (DFG) project nr. 415510302.
\end{acknowledgements}

\bibliographystyle{spbasic}      
\bibliography{joined}
\end{document}